\documentclass[twocolumn]{aastex631}
\usepackage{amsmath}
\usepackage{longtable}
\begin{document}

\title{Hydrodynamics and Nucleosynthesis of Jet-Driven Supernovae II: Comparisons with Abundances of Extremely Metal-Poor Galaxies and Constraints on Supernova Progenitors}

\shortauthors{Leung and Nomoto}
\shorttitle{Galactic Chemical Enrichment by Jet-Driven Supernovae}

\author[0000-0002-4972-3803]{Shing-Chi Leung}

\affiliation{Department of Mathematics and Physics, SUNY Polytechnic Institute, 100 Seymour Road, Utica, NY 13502, USA}

\author[0000-0001-9553-0685]{Ken'ichi Nomoto}

\affiliation{Kavli Institute for the Physics and 
Mathematics of the Universe (WPI), The University 
of Tokyo Institutes for Advanced Study, The 
University of Tokyo, Kashiwa, Chiba 277-8583, Japan}

\correspondingauthor{Shing-Chi Leung}
\email{leungs@sunypoly.edu}

\date{\today}

\submitjournal{ApJ on Dec 21 2023}
\received{Dec 21 2023}
\revised{Aug 3 2024}
\accepted{Aug 8 2024}
\published{Oct 18 2024}

\newcommand{\red}[1]{\textcolor{red}{#1}}

\begin{abstract}

The spectra of several galaxies, including extremely metal-poor galaxies (EMPGs) from the EMPRESS survey, have shown that the abundances of some Si-group elements differ from ``spherical'' explosion models of massive stars. This leads to the speculation that these galaxies have experienced supernova explosions with high asphericity, where mixing and fallback of the inner ejecta with the outer material leads to the distinctive chemical compositions. In this article, we consider the jet-driven supernova models by direct two-dimensional hydrodynamics simulations using progenitors about 20 -- 25 $M_{\odot}$ at zero metallicity. We investigate how the abundance patterns depend on the progenitor mass, mass cut and the asphericity of the explosion. We compare the observable with available supernova and galaxy catalogs based on $^{56}$Ni, ejecta mass, and individual element ratios. The proximity of our results with the observational data signifies the importance of aspherical supernova explosions in chemical evolution of these galaxies.  Our models will provide the theoretical counterpart for understanding the chemical abundances of high-z galaxies measured by the James Webb Space Telescope.

\end{abstract}





\section{Introduction}

\subsection{Aspherical Supernova Explosion}

Many observable features of core collapse supernovae (CCSNe) suggest that the their explosion could be aspherical. The clumpiness, filament structures, and the observed directional variations of the ejecta imply an aphserical explosion histories. It is further supported by evidences including (1) the ring-shape structure of the SN 1987A \citep{Wang1992SN1987A} and the clear bipolar motion of high-speed knots in opposite directions \citep{Fesen2001}, (2) the level of polarization in SN 1993J \citep{Tran1997} and SN 1998bw \citep{Hoefflich1999, Dessart2017SN1998bw}, (3) the high speed radioactive "bullets" in the Vela supernova remnant \citep{Strom1995Vela}, and (4) the high velocity compact remnants discovered in the Vela supernova \citep{Strom1995Vela}.

The aspherical explosion can be resulted from standard CCSNe by the neutrino-driven explosions through the Standing Accretion Shock Instability \citep{Foglizzo2015SASI}. The recent modeling approaches and results are reviewed in \cite{Janka2016CCSNReview}, \cite{Burrows2021CCSNReview}, and \cite{Mueller2020}. However, the actual explosion is subject to the intricate input physics, such as the equation of states, the neutrino physics and the progenitors \citep[e.g.,][]{Burrows2018CCSNReview}. Variations such as the magneto-rotational explosion and the outburst from the black hole accretion disk (i.e., jet-driven explosion or collapsar) are also possible candidates for explaining aspherical explosions. 

Simulations of jet-driven explosions from an accretion disk wind have first been suggested by \cite{Khokhlov1999Collapsar}. How the jet propagates through the stellar envelope is studied extensively in e.g., \cite{MacFadyen2001Collapsar}. The models are applied to explain the polarimetric observations of some Type II SNe \citep{Couch2009} and some shock-breakout light curves such as SN 2008D \citep{Couch2011SN2008D}.

The magneto-rotational instability \citep[e.g.,][]{LeBlanc1970Magnetar, Obergaulinger2009MSI, Obergaulinger2023CollapsarReview, Mueller2024} is required for a jet-like structure to develop outside a neutron star. But recent 3D simulations suggested that the collimation could be limited by kink instabilities \citep{Moesta2014, Kuroda2020}. In a similar background, the magnetic field can power the jet from the black hole accretion disk \citep{Fujimoto2006AccretionDisk}. Similarly, neutrino also affects the disk dynamics and its nucleosynthesis \citep{Surman2004}. In both cases, the rotation is a key for the accretion disk to grow, which can lead to a jet outburst which supplies a tremendous amount of energy ($\sim 10^{51}$ erg) to the envelope. The rotation model is particularly important because massive stars with a mass $\geqslant 18~M_{\odot}$ might explode as a hypernova\footnote{We remind that the term hypernova is used in the spherical explosion models which require more than 10 times higher of the explosion energies than normal bright SNe. It outlines the difference for supernova models with or without GRBs \citep{Iwamoto1998Hypernova}.} 
to explain the bifurcation of explosion events \citep{Nomoto2003HN} .

For rotating models, how fast the star rotates depends on the angular momentum loss, which is hinged on the stellar mass loss. In metal-poor stars, the lack of metals for line absorption suppresses the stellar mass loss rate and a massive H-envelope persists \citep[e.g., ][]{Marigo2001, Martinet2023RotatingFirstStar, Song2020RotatingStar}. Depending on the transfer of angular momentum from the core to the envelope, the core is suggested to obtain a high rotation velocity \citep{Murphy2021}, which favours the aspherical explosions. Some rotating first star models studied in \cite{Ekstrom2008RotatingFirstStars} show Fe cores capable of forming a maximally rotating black holes. These black holes will be the potential site for the later jet outburst.

Once the jet forms, the cone-shaped energy deposition creates a very different thermodynamical history from a spherical explosion. This creates enhanced production around $A = 45$ and 65 \citep[e.g., ][]{Nagataki1997}. The high temperature creates the alpha-rich freezeout, which is essential for producing individual isotopes such as $^{44}$Ca, $^{47-48}$Ti, $^{59}$Co \citep[e.g.,][]{Maeda2002Hypernovae}. Similar results are observed for the zero-metallicity star explosion performed in \cite{Tominaga2009}. The jet energy deposition triggers aspherical mass ejection, which can explain the carbon-enriched metal-poor stars (CEMPs) \citep{Tominaga2007jet}, and some Zn-rich stars such as HE 1327-2326 \citep{Ezzeddine2019}. The jet structure also favours the ejection of $^{56}$Ni-rich matter with a high velocity \citep{Nagataki1998, Mazzali2005, Maeda2006}, which has been observed in some supernovae and hypernovae (e.g. SN 1987A, SN 1998bw). The energy composition of the jet also leaves signature by the polarization of the ejecta \citep[e.g.,][]{Couch2009}. 

\subsection{Spectroscopic Observations}

The search for high-redshift galaxies has opened another dimension in understanding the nature of supernova explosions, especially low metallcity ones. Similar to extremely metal-poor stars, extremely metal-poor galaxies (EMPGs) are ideal candidates to capture the effects of early massive supernova explosion by their chemical composition, which reflects the explosive yield of one or a few supernova explosions \citep[e.g.,][]{Hartwig2018}. Such early galaxies also imply that the Type Ia supernovae are less important, which have a delay time from the formation of white dwarfs to their final explosions \citep[e.g.,][]{Kobayashi2020}. Thus the composition of these extreme metal-poor objects provides direct constraints on the massive star explosion yields. Recent surveys, and their follow-up observations, have discovered several galactic objects with unusual Si- and Fe-group elements, including Ne, Si, S, Ar, Fe, and Ni. 

In \cite{Kojima2020} the Extremely Metal-Poor Representatives Explored by the Subaru Survey (EMPRESS) is designed to use the Subaru/Hyper Supreme-Cam (HSC) optical images to identify the faint EMPGs. Among the $>100$ EMPGs identified, their follow-up project later selected 13 galaxies and studied them in details, where the galaxies have a metallicity $Z \sim 0.1-1\% Z_{\odot}$ \citep{Isobe2022}. They measured the $\alpha$-chain elements O, Ne, Ar and Fe, which can be directly connected to the massive star progenitors and explosion mechanisms. Among all features, these galaxies have the [Ne/O], [Ar/O] and [Fe/O] values very close to solar values or mildly sub-solar, despite their high metal-poor nature. 
 


In \cite{Watanabe2023}, the 13$^{\rm th}$ follow-up project on the analysis of the galaxy catalogue from EMPRESS project is reported for more EMPGs, including SBS-0335-052E \citep{Izotov2018}, J2314+0154 \citep{Kojima2020}, and J0125+0759 \citep{Kojima2020}. The spectra of these galaxies are measured by Keck/LRIS. Ratios of  [Ne/O], [Ar/O], [S/O] and [Fe/O] are reported. In that work, the mixing-fallback model is extensively adopted to explain the unconventional abundance pattern. 
These galaxies are shown to fit with a small mixing ratio, about 0 -- 0.2, however, the [Fe/O] cannot be fully explained by the mixing and fallback mechanisms. [Ar/O] and [S/O] could be indicators of PISN explosions.

\subsection{Motivation}

The chemical abundances provide first hand constraints on how massive stars explode and eject matter to the surrounding \citep[e.g.,][]{Thielemann2019Nucleo, Arcones2023SNReview}. The discrepancies between the measured galaxies' chemical abundance and nucleosynthesis yields of ordinary stars have led to comparisons with less canonical supernova models, including hypernovae and pair-instability supernova models \citep{Isobe2022}. While the early galaxies favour the formation of very massive stars \citep[e.g.,][]{Hirano2015}, later radiation hydrodynamics simulations suggest ordinary massive stars \citep[see e.g.,][]{Latif2022}. This implies that both ordinary and very massive stars are possible candidates for the chemical origin of these galaxies. The mismatch of chemical abundances of these galaxies with nucleosynthetsis yields of spherical models leads to the question, if other models can explain the abundance pattern. For example, very massive stars \citep{Woods2020VeryMassiveStar, Umeda2024} are often invoked for matching less commonly seen nucleosynthetic patterns.

In one-dimensional simulations, the mixing-fallback model has been invoked to mimic the multi-dimensional mixing process \citep{Umeda2002}. The mixing-fallback approximation has successfully explained the peculiar abundance patterns of some metal-poor stars, such as the very high C/Fe abundance ratio \citep[e.g.,][]{Umeda2002,Ishigaki2018}. The mixing processes have been shown in neutrino-driven explosions as exhibited by their highly aspherical structure \cite[see characteristic simulations from various groups, e.g.,][]{Burrows20123DCCSN, Melson2015, Lentz2015CCSN15}.

In Paper I \citep{Leung2023Jet1}, we presented a catalogue of the jet-driven supernova models using multi-dimensional simulations. The model uses the 40 $M_{\odot}$ zero metallicity pop III star as the progenitor. The jet-like thermal bomb is applied to parametrize the explosion process in the center. We studied nucleosynthesis of the jet-triggered explosion and showed how the chemical composition depends on the jet energetics. The yields of the jet models exhibit significant differences from spherical explosion models. The jet explosion is essential in reproducing the trend of some element pair such as Ti and V. It is therefore interesting to extend our models, and check if the results derived from a more consistent modeling provide better fitting. 
 
In this article, we first review the essential massive star models presented in this work in Section \ref{sec:models}. We describe how the massive star explosion models are prepared in both one- and two-dimensions. Then in Section \ref{sec:results} we present how the hydrodynamics and nucleosynthesis depend on the dimensionality, progenitor mass and other input physics. In Section \ref{sec:discussion} we analyze the elemental trends of some key element pairs, and compare them with the observed metal-poor stars and supernovae. We also present the elemental yields of our models. We apply our results to compare with some well-observed metal-poor galaxies, and some metal-poor stars in Section \ref{sec:mps}. 
At last we give our conclusion. 

This article uses the convention
\begin{equation}
    [\textrm{X/Fe}] = \log_{10} \frac{(\rm{X/Fe})}{(\rm{X/Fe})_{\odot}}.
\end{equation}

\section{Massive Star Models}
\label{sec:models}


\subsection{Two-dimensional Simulations}

In Paper I \citep{Leung2023Jet1}, we studied the dependence of explosive nucleosynthesis on the jet explosion parameters for the 40 $M_{\odot}$ star progenitor using the special relativistic hydrodynamics solver extended from \cite{Leung2015Code}. The jet energy deposition rate, its duration, and the open angles are treated as main parameters. We followed \cite{Tominaga2009} and set the {\sl standard} jet as $\dot{E}_{\rm dep,0} = 1.2 \times 10^{53}$ ergs$^{-1}$, $E_{\rm dep,0} = 1.5 \times 10^{52}$ erg and a jet open angle of 15$^{\circ}$. This implies a jet deposition time of $t_{\rm jet,0} = 0.125$ s.

In each model, we first remove the core material interior to the {\sl mass cut}, and then we deposit the energy at the inner boundary of the jet with the jet prescription above. We adopt the Helmholtz equation of state \citep{Timmes1999Helm, Timmes2000Helm} and a simple 7-isotope network
\citep{Timmes20007iso} to describe nuclear reactions in the hydrodynamics simulations. We calculate the explosion until most exothermic nuclear reactions have terminated and the ejecta appears to be in homologous expansion to a good approximation. 

In the simulations, we use the tracer particle scheme \citep[e.g., ][]{Arnett1989SN1987A, Hachisu1990, Travaglio2004, Seitenzahl2010Tracer} to record the thermodynamic history of the underlying fluid motion. After the hydrodynamics simulations, we use the thermodynamic histories of the tracer particles to reconstruct the detailed nucleosynthesis with the 495-isotope network containing isotopes from H to Tc \citep{Timmes1999Torch}.

We extend our calculation to include progenitors with the zero-age main-sequence (ZAMS) masses of $M_{\rm ZAMS} = $ 20 and 25 $M_{\odot}$ presented in \cite{Tominaga2007}. Recent pre-supernova and core-collapse models \citep[e.g., ][]{Sukhbold2016} have demonstrated the possibilities of the formation of an accretion disk and a following outburst. 
This also agrees with the suggestions by \cite{Nomoto2013} that the final fate of stars with $M_{\rm ZAMS} \geqslant 25~M_{\odot}$ can bifurcate into faint supernovae and hypernovae following fallback. In Tables \ref{table:S20models}-\ref{table:S25cmodels} we list the new models computed in this work and some of their essential parameters. For convenience we name all the models in the following format: a S20-0500-2000-15 corresponds to the progenitor model with $M_{\rm ZAMS} = 20 M_{\odot}$ (S20), the jet with the energy deposition rate of 0.5 $\dot{E}_{\rm jet,0}$ (0500), jet deposition time of 2.0 $t_{\rm jet,0}$ (2000), an open angle of $15^{\circ}$ (15). All models assume the inner boundary at 900 km, except for the series S25b (s25c) corresponds to the model sequences where the innermost boundary is located at 1500 (2100) km. In Appendix A we further tabulate the elemental ratios of other major elements from these models. 

\begin{table*}[]
    \caption{The jet-driven supernova models using the 20 $M_{\odot}$ ZAMS star for the progenitor. $M_{\rm ZAMS}$, $M_{\rm pro}$, $M_{\rm ej}$, and $M$(O) are the progenitor ZAMS mass, pre-collapse, ejecta and ejected O- masses in units of $M_{\odot}$. $\dot{E}_{\rm dep}$, $t_{\rm dep}$, and $E_{\rm dep}$ are the energy deposition rate, deposition time and total deposited energy in units of the characteristic model. [Ne/O], [Si/O], [S/O], [Ar/O] and [Fe/O] are the ejecta mass fraction ratios of the element pairs, defined in Equation (1). The model name SWW-XXXX-YYYY-ZZ stands for the model using ZAMS mass WW $M_{\odot}$, XXXX$/1000$ $\dot{E}_{\rm ref}$, YYYY$/1000$ $t_{\rm ref}$, and an ZZ$^{\circ}$ open angle. See also Table \ref{table:yields2_S20} for C-Zn of this model series.} 
    \centering
    \begin{tabular}{c c c c c     c c c c c    c c c c}
        \hline
         Model & $M_{\rm ZAMS}$ & $M_{\rm pro}$ & $\dot{E}_{\rm dep}$ & $t_{\rm dep}$ & $E_{\rm dep}$ & $\theta_{\rm jet}$ & $M_{\rm ej}$ & M(O) & [Ne/O] & [Si/O] & [S/O] & [Ar/O] & [Fe/O] \\ \hline
         S20-0250-1000-15 & 20 & 4.08 & 0.25 & 1.00 & 0.375 & 15 & 1.94 & 1.02 & 0.14 & -1.01 & -1.08 & -1.17 & -0.70 \\
         S20-0500-0500-15 & 20 & 4.08 & 0.50 & 0.50 & 0.375 & 15 & 2.62 & 1.42 & 0.14 & -0.80 & -0.88 & -0.97 & -0.85 \\
         S20-0500-1000-15 & 20 & 4.08 & 0.50 & 1.00 & 0.750 & 15 & 2.81 & 1.53 & 0.14 & -0.73 & -0.77 & -0.86 & -0.79 \\
         S20-0500-2000-15 & 20 & 4.08 & 0.50 & 2.00 & 1.500 & 15 & 2.91 & 1.59 & 0.13 & -0.71 & -0.75 & -0.83 & -0.80 \\
         S20-1000-0500-15 & 20 & 4.08 & 1.00 & 0.50 & 0.750 & 15 & 3.17 & 1.58 & 0.13 & -0.64 & -0.66 & -0.72 & -0.83 \\
         S20-1000-0250-15 & 20 & 4.08 & 1.00 & 0.25 & 0.375 & 15 & 2.87 & 1.73 & 0.14 & -0.82 & -0.94 & -1.05 & -0.90 \\
         S20-1000-1000-15 & 20 & 4.08 & 1.00 & 1.00 & 1.500 & 15 & 3.18 & 1.76 & 0.13 & -0.65 & -0.68 & -0.76 & -1.05 \\
         S20-1000-2000-15 & 20 & 4.08 & 1.00 & 2.00 & 3.000 & 15 & 3.25 & 1.81 & 0.13 & -0.60 & -0.64 & -0.71 & -1.05 \\
         S20-1000-4000-15 & 20 & 4.08 & 1.00 & 4.00 & 6.000 & 15 & 3.34 & 1.78 & 0.13 & -0.61 & -0.63 & -0.70 & -0.84 \\
         S20-2000-0500-15 & 20 & 4.08 & 2.00 & 0.50 & 1.500 & 15 & 3.48 & 1.81 & 0.13 & -0.52 & -0.51 & -0.57 & -0.76 \\
         S20-2000-1000-15 & 20 & 4.08 & 2.00 & 1.00 & 3.000 & 15 & 3.46 & 1.91 & 0.12 & -0.50 & -0.51 & -0.59 & -0.87 \\
         S20-2000-2000-15 & 20 & 4.08 & 2.00 & 2.00 & 6.000 & 15 & 3.55 & 1.88 & 0.11 & -0.40 & -0.38 & -0.43 & -0.82 \\
         S20-4000-1000-15 & 20 & 4.08 & 4.00 & 1.00 & 6.000 & 15 & 3.59 & 1.94 & 0.12 & -0.34 & -0.32 & -0.37 & -0.74 \\
         S20-4000-2000-15 & 20 & 4.08 & 4.00 & 2.00 & 12.000 & 15 & 3.56 & 1.94 & 0.12 & -0.40 & -0.40 & -0.47 & -0.79  \\
        \hline         
    \end{tabular}
    \label{table:S20models}
\end{table*}


\begin{table*}[]
    \caption{Same as Table \ref{table:S20models} but for models using the 25 $M_{\odot}$ progenitor as the pre-collapse model. See also Table \ref{table:yields2_S25a} for C-Zn of this model series.}
    \centering
    \begin{tabular}{c c c c c     c c c c c   c c c c}
        \hline
         Model & $M_{\rm ZAMS}$ & $M_{\rm pro}$ & $\dot{E}_{\rm dep}$ & $t_{\rm dep}$ & $E_{\rm dep}$ & $\theta_{\rm jet}$ & $M_{\rm ej}$ & $M$(O) & [Ne/O] & [Si/O] & [S/O] & [Ar/O] & [Fe/O] \\ \hline
         S25-0250-1000-15 & 25 & 5.58 & 0.25 & 1.00 & 0.375 & 15 & 1.76 & 0.64 & -0.33 & -0.11 & -0.01 & -0.06 & -0.02 \\
         S25-0250-2000-15 & 25 & 5.58 & 0.25 & 2.00 & 0.750 & 15 & 1.51 & 0.61 & -0.41 & -0.06 & 0.02 & -0.03 & -0.89  \\
         S25-0500-0500-15 & 25 & 5.58 & 0.50 & 0.50 & 0.375 & 15 & 2.62 & 1.07 & -0.30 & -0.33 & -0.26 & -0.35 & -0.29 \\
         S25-0500-1000-15 & 25 & 5.58 & 0.50 & 1.00 & 0.750 & 15 & 2.74 & 1.26 & -0.35 & -0.14 & -0.07 & -0.13 & -0.19 \\
         S25-0500-2000-15 & 25 & 5.58 & 0.50 & 2.00 & 1.500 & 15 & 2.50 & 1.22 & -0.36 & -0.20 & -0.14 & -0.20 & -0.57 \\
         S25-0500-4000-15 & 25 & 5.58 & 0.50 & 4.00 & 3.000 & 15 & 3.02 & 1.44 & -0.34 & -0.14 & -0.08 & -0.15 & -0.20 \\
         S25-1000-0250-15 & 25 & 5.58 & 1.00 & 0.25 & 0.375 & 15 & 1.50 & 0.60 & -0.33 & -0.62 & -0.60 & -0.67 & -0.41  \\
         S25-1000-0500-15 & 25 & 5.58 & 1.00 & 0.50 & 0.750 & 15 & 3.31 & 1.67 & -0.31 & -0.20 & -0.16 & -0.22 & -0.31  \\
         S25-1000-1000-15 & 25 & 5.58 & 1.00 & 1.00 & 1.500 & 15 & 3.61 & 1.78 & -0.33 & -0.12 & -0.08 & -0.13 & -0.20 \\
         S25-1000-1000-30 & 25 & 5.58 & 1.00 & 1.00 & 1.500 & 30 & 3.75 & 1.84 & -0.37 & -0.08 & -0.05 & -0.11 & -0.11 \\
         S25-1000-2000-15 & 25 & 5.58 & 1.00 & 2.00 & 3.000 & 15 & 3.62 & 1.80 & -0.32 & -0.13 & -0.09 & -0.16 & -0.19 \\
         S25-1000-2000-30 & 25 & 5.58 & 1.00 & 2.00 & 3.000 & 30 & 3.84 & 1.87 & -0.37 & -0.04 & -0.01 & -0.06 & 0.01 \\
         S25-1000-4000-15 & 25 & 5.58 & 1.00 & 4.00 & 6.000 & 15 & 3.20 & 1.74 & -0.32 & -0.26 & -0.26 & -0.34 & -0.75 \\
         S25-2000-0250-15 & 25 & 5.58 & 2.00 & 0.25 & 0.750 & 15 & 3.63 & 1.84 & -0.29 & -0.27 & -0.26 & -0.32 & -0.19 \\
         S25-2000-0500-15 & 25 & 5.58 & 2.00 & 0.50 & 1.500 & 15 & 3.97 & 1.99 & -0.34 & -0.12 & -0.11 & -0.17 & -0.17 \\
         S25-2000-1000-15 & 25 & 5.58 & 2.00 & 1.00 & 3.000 & 15 & 4.00 & 2.04 & -0.36 & -0.11 & -0.11 & -0.18 & -0.20 \\
         S25-2000-1000-30 & 25 & 5.58 & 2.00 & 1.00 & 3.000 & 30 & 4.12 & 2.02 & -0.39 & -0.03 & -0.01 & -0.07 & -0.00 \\
         S25-2000-2000-15 & 25 & 5.58 & 2.00 & 2.00 & 6.000 & 15 & 4.02 & 2.09 & -0.41 & -0.08 & -0.11 & -0.19 & -0.29 \\
         S25-2000-2000-30 & 25 & 5.58 & 2.00 & 2.00 & 6.000 & 30 & 4.23 & 2.01 & -0.43 & 0.04 & 0.07 & 0.01 & 0.07 \\
         S25-4000-0500-15 & 25 & 5.58 & 4.00 & 0.50 & 3.000 & 15 & 4.33 & 2.17 & -0.41 & -0.01 & -0.00 & -0.06 & -0.15 \\
         S25-4000-1000-15 & 25 & 5.58 & 4.00 & 1.00 & 6.000 & 15 & 4.29 & 2.14 & -0.49 & 0.09 & 0.11 & 0.05 & -0.19  \\
         S25-4000-2000-15 & 25 & 5.58 & 4.00 & 2.00 & 12.00 & 15 & 4.44 & 2.08 & -0.36 & -0.11 & -0.15 & -0.23 & -0.14 \\
        \hline         
    \end{tabular}
    \label{table:S25models}
\end{table*}

\begin{figure*}
    \centering
    \includegraphics[width=8.5cm]{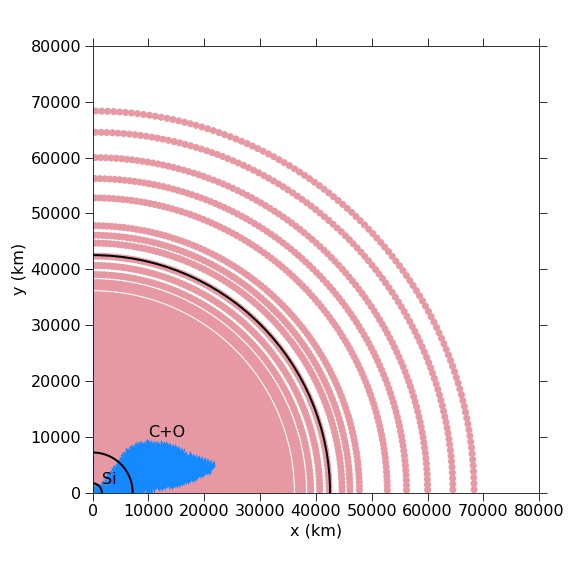}
    \includegraphics[width=8.5cm]{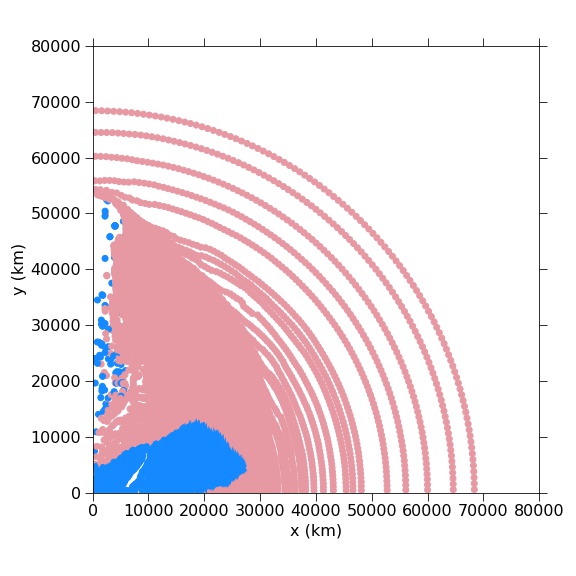}
    \includegraphics[width=8.5cm]{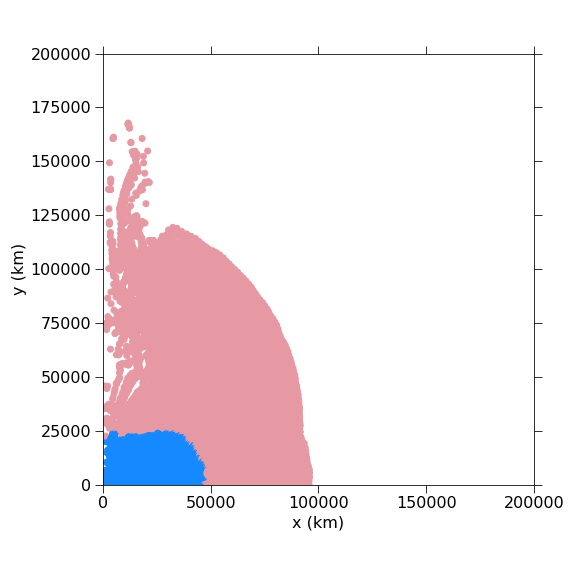}
    \includegraphics[width=8.5cm]{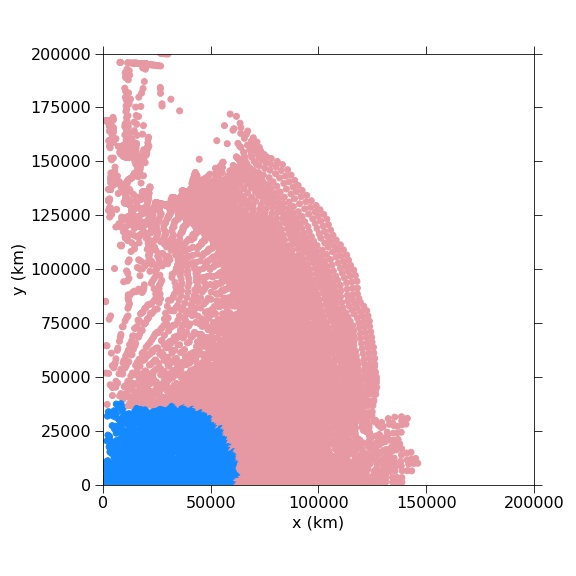}
    \caption{The tracer distribution in the characteristic 25 $M_{\odot}$ model, with red being ejected and blue being bound. The Si and C+O cores are indicated by the solid line for illustration. The top left, top right, bottom left and bottom right panels stand for four snapshots at equal time steps of $\sim 4$ s, starting from the initial profile.}
    \label{fig:N25_tracers}
\end{figure*}

 In Figure \ref{fig:N25_tracers} we plot the tracer distribution of the characteristic model using the 25 $M_{\odot}$ star progenitor. We use the final energy of the tracers to determine if they are ejected or bound. 
 The Si layer within the cone shape is ejected. The inner C+O layer has an increasing spread in the ejecta angular range until $\sim 20,000$ km. Beyond that, the outer shell is ejected. 

\begin{figure}
    \centering
    \includegraphics[width=8.5cm]{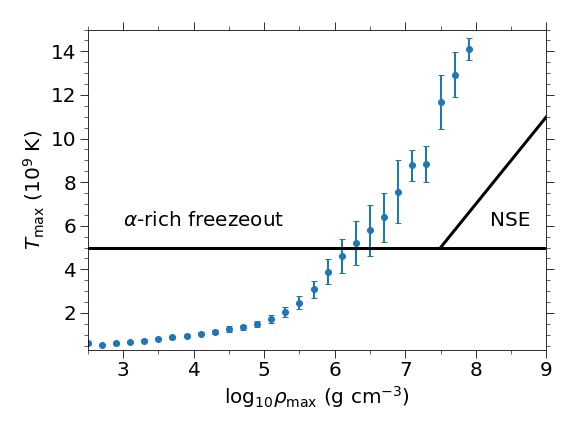}
    \caption{Statistics of the tracers for the peak temperature $T_{\rm max}$ and their corresponding density  in the characteristic model S25-1000-1000-15.}
    \label{fig:tracer_benchmark_S25}
\end{figure}

We summarize the tracer thermodynamic history in Figure \ref{fig:tracer_benchmark_S25}. It collects the statistics of the passive tracers in the characteristic model when the tracers reach their maximum temperature. We bin the tracers according to their corresponding density. In Figure \ref{fig:tracer_benchmark_S25}, the data points correspond to the average of the tracers within individual density bins and the error bars stand for the standard deviation of the data. 

The thermodynamics provides a useful diagnosis to the nuclear reaction class where the tracers have experienced. In general, high density and temperature burning leads to nuclear statistical equilibrium (NSE), where the composition can be solely determined by $\rho, T$, and $Y_{\rm e}$. In NSE, the final composition can be predicted by the density where the matter leaves NSE. On the other hand, for a low density and high temperature environment, $\alpha$-burning becomes the bottleneck of the entire nuclear reaction network. Thus, not all isotopes are equally accessible as in NSE. This gives rise to a unique chemical abundance pattern featuring high entropy isotopes, e.g., $^{64}$Zn. 

In the characteristic model of the 25 $M_{\odot}$ star, we observe an almost monotonically increasing trend for the tracers. Tracers in the inner part of the star, having higher densities, experienced higher peak temperatures in their burning history. There are observable fluctuations for tracers from the inner ejecta. It is because the ejecta includes both the region directly excited by the jet, and the nearby region which follows the expansion. Matter in the outer region is mostly ejected without experiencing a significant shock heating. The spherical-like mass ejection agrees with the low fluctuations among tracers.

\begin{table*}[]
    \caption{Same as Table \ref{table:S25models} but with an innermost boundary at 1500 km. See also Table \ref{table:yields2_S25b} for C-Zn of this model series.}
    \centering
    \begin{tabular}{c c c c c     c c c c c    c c c c}
        \hline
         Model & $M_{\rm ZAMS}$ & $M_{\rm pro}$ & $\dot{E}_{\rm dep}$ & $t_{\rm dep}$ & $E_{\rm dep}$ & $\theta_{\rm jet}$ & $M_{\rm ej}$ & $M$(O) & [Ne/O] & [Si/O] & [S/O] & [Ar/O] & [Fe/O] \\ \hline
         S25b-0500-1000-15 & 25 & 5.34 & 0.50 & 1.00 & 0.750 & 15 & 1.55 & 0.61 & -0.33 & -0.13 & -0.04 & -0.10 & -0.14 \\
         S25b-0500-2000-15 & 25 & 5.34 & 0.50 & 2.00 & 1.500 & 15 & 1.41 & 0.60 & -0.37 & -0.22 & -0.12 & -0.16 & -0.67 \\
         S25b-1000-0500-15 & 25 & 5.34 & 1.00 & 0.50 & 0.750 & 15 & 2.06 & 1.00 & -0.35 & -0.22 & -0.15 & -0.20 & -0.50 \\
         S25b-1000-1000-15 & 25 & 5.34 & 1.00 & 1.00 & 1.500 & 15 & 2.37 & 1.14 & -0.34 & -0.18 & -0.12 & -0.17 & -0.29 \\
         S25b-1000-2000-15 & 25 & 5.34 & 1.00 & 2.00 & 3.000 & 15 & 2.46 & 1.19 & -0.34 & -0.15 & -0.10 & -0.16 & -0.26  \\
         S25b-1000-4000-15 & 25 & 5.34 & 1.00 & 4.00 & 6.000 & 15 & 2.48 & 1.34 & -0.36 & -0.24 & -0.25 & -0.33 & -0.76 \\
         S25b-2000-0500-15 & 25 & 5.34 & 2.00 & 0.50 & 1.500 & 15 & 2.83 & 1.43 & -0.33 & -0.16 & -0.13 & -0.19 & -0.30 \\
         S25b-2000-1000-15 & 25 & 5.34 & 2.00 & 1.00 & 3.000 & 15 & 2.91 & 1.47 & -0.35 & -0.12 & -0.10 & -0.16 & -0.29 \\
         S25b-2000-2000-15 & 25 & 5.34 & 2.00 & 2.00 & 6.000 & 15 & 2.92 & 1.49 & -0.36 & -0.13 & -0.14 & -0.21 & -0.28 \\
         S25b-4000-1000-15 & 25 & 5.34 & 4.00 & 1.00 & 6.000 & 15 & 3.03 & 1.59 & -0.40 & -0.00 & 0.03 & -0.03 & -0.39 \\
         S25b-4000-2000-15 & 25 & 5.34 & 4.00 & 2.00 & 12.000 & 15 & 3.06 & 1.58 & -0.47 & 0.10 & 0.14 & 0.08 & -0.42 \\
        \hline         
    \end{tabular}
    \label{table:S25bmodels}
\end{table*}

\begin{table*}[]
    \caption{Same as Table \ref{table:S25models} but with an innermost boundary at 2100 km. See also Table \ref{table:yields2_S25c} for C-Zn of this model series.}
    \centering
    \begin{tabular}{c c c c c     c c c c c   c c c c}
        \hline
         Model & $M_{\rm ZAMS}$ & $M_{\rm pro}$ & $\dot{E}_{\rm dep}$ & $t_{\rm dep}$ & $E_{\rm dep}$ & $\theta_{\rm jet}$ & $M_{\rm ej}$ & $M$(O) & [Ne/O] & [Si/O] & [S/O] & [Ar/O] & [Fe/O] \\ \hline
         S25c-0500-1000-15 & 25 & 5.21 & 0.50 & 1.00 & 0.750 & 15 & 1.48 & 0.52 & -0.29 & -0.31 & -0.25 & -0.32 & -0.13 \\
         S25c-0500-2000-15 & 25 & 5.21 & 0.50 & 2.00 & 1.500 & 15 & 2.00 & 0.77 & -0.33 & -0.14 & -0.05 & -0.12 & -0.35 \\
         S25c-1000-0500-15 & 25 & 5.21 & 1.00 & 0.50 & 0.750 & 15 & 2.05 & 0.78 & -0.34 & -0.21 & -0.13 & -0.18 & -0.23 \\
         S25c-1000-1000-15 & 25 & 5.21 & 1.00 & 1.00 & 1.500 & 15 & 2.34 & 0.98 & -0.33 & -0.21 & -0.14 & -0.20 & -0.36 \\
         S25c-1000-2000-15 & 25 & 5.21 & 1.00 & 2.00 & 3.000 & 15 & 2.90 & 1.41 & -0.34 & -0.23 & -0.19 & -0.27 & -0.59 \\
         S25c-1000-4000-15 & 25 & 5.21 & 1.00 & 4.00 & 6.000 & 15 & 3.40 & 1.78 & -0.35 & -0.22 & -0.21 & -0.28 & -0.74  \\
         S25c-2000-0500-15 & 25 & 5.21 & 2.00 & 0.50 & 1.500 & 15 & 3.15 & 1.52 & -0.33 & -0.19 & -0.14 & -0.21 & -0.40 \\
         S25c-2000-1000-15 & 25 & 5.21 & 2.00 & 1.00 & 3.000 & 15 & 3.63 & 1.83 & -0.32 & -0.16 & -0.12 & -0.18 & -0.45 \\
         S25c-2000-2000-15 & 25 & 5.21 & 2.00 & 2.00 & 6.000 & 15 & 4.00 & 2.06 & -0.37 & -0.08 & -0.07 & -0.15 & -0.42 \\
         S25c-4000-1000-15 & 25 & 5.21 & 4.00 & 1.00 & 6.000 & 15 & 3.83 & 2.06 & -0.35 & -0.12 & -0.10 & -0.16 & -0.70 \\
         S25c-4000-2000-15 & 25 & 5.21 & 4.00 & 2.00 & 12.000 & 15 & 4.02 & 2.11 & -0.43 & 0.01 & 0.02 & -0.05 & -0.48  \\
        \hline         
    \end{tabular}
    \label{table:S25cmodels}
\end{table*}

\begin{figure}
    \centering
    \includegraphics[width=8.5cm]{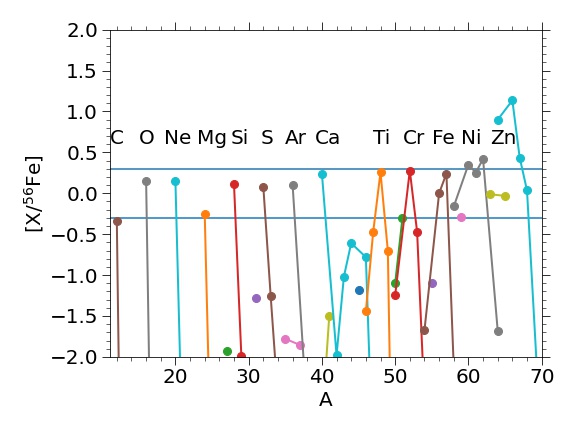}
    \caption{The scaled isotope fraction ($[X/$Fe$] = \log_{10} (X/$Fe)$/(X/$Fe$)_{\odot}$) for elements from C to Zn. 
    The two horizontal lines stand for 50\% and 200\% of the solar value. 
    }
    \label{fig:xiso_benchmark_N25}
\end{figure}

In Figure \ref{fig:xiso_benchmark_N25} we plot the characteristic model using the 25 $M_{\odot}$ progenitor.
The two-dimensional model shows a flat distribution of chemical elements from C to Zn. Most elements along the $\alpha$-chain have abundances compatible with the solar abundance. Odd number elements such as P, K, Sc, and the low-Ye isotopes (i.e., the isotopes with a higher neutron number) are mostly underproduced. The zero-metallicity environment does not contain any $^{22}$Ne to directly produce the low-Ye isotopes, especially in the Si-group elements. 

\subsection{Mixing-Fallback Mechanisms}

The Mixing-Fallback model \citep{Umeda2002} is an approximate modeling of the jet-like aspherical mass ejection by using the spherical explosion model. 
The mixing-fallback model can reproduce the abundance pattern of the jet-like explosion rather well \citep{Tominaga2009}. The model considered the process that, 
instead of the ejection of the entire star, a part of the inner core and outer envelope are mixed and ejected, and the remaining matter falls back and accreted onto the central compact object. The model assumes three parameters: (1) the inner mass-cut $M_{\rm cut,~in}$, (2) the outer mass-cut $M_{\rm cut,~out}$ and (3) the mixing ratio $f_{\rm mix}$. The mass interior to the inner mass-cut all accretes. The total accreted mass is given by
\begin{equation}
    M_{\rm acc} = M_{\rm cut,in} + (1 - f_{\rm mix}) (M_{\rm cut,out} - M_{\rm cut,in})
\end{equation}
The exact mass of individual elements then depends on both the composition of the middle layer and the outer layer. This model has been used extensively to describe the origin of the peculiar chemical abundance patterns in extremely metal-poor stars \citep[e.g.,][]{Iwamoto2005,Ishigaki2018}. 

The mixing-fallback model can reproduce the abundance pattern of jet-like explosion rather well \cite{Tominaga2009}. However, the spherical model is different from a full 2- or 3D models. Assuming the explosion of the same energy, the energy is deposited in a spherical shell, rather than being focused on the cone shaped structure. Thus, the mixing-fallback model has (1) a less active nucleosynthesis and (2) a lower ejecta velocity. In Appendix B, we tabulate the stable and short-lived radioactive isotopes of the spherical models with various masses and explosion energy, together with their elemental production. 

\section{Explosive Observable and Nucleosynthesis}
\label{sec:results}

\subsection{Dependence on Dimensionality}

The most important effect of the jet-driven explosion on supernova models is that the jet provides local heating for the matter to expand and to be ejected. This cannot be achieved in classical spherical models. In the aspherical model compared with the spherical model, a small explosion energy is enough for the ejecta to reach such high velocities as observed in some hypernovae or highly aspherical supernovae \citep[e.g., SN 1998bw studied in][]{Nakamura2001, Hoefflich1999, Mazzali2001}. This allows the aspherical model to synthesize and eject Fe-group elements more readily than the spherical model with the same explosion energy. Here we examine how the dimensionality changes nucleosynthetic products.

\begin{figure}
    \centering
    \includegraphics[width=8.5cm]{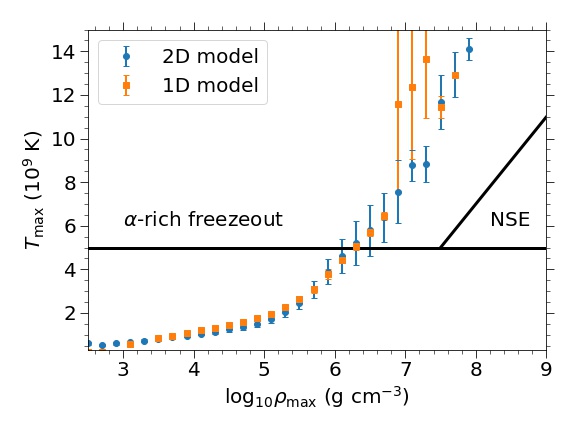}
    \caption{The tracer thermodynamics history for the 25 $M_{\odot}$ star assuming spherical (1D: orange squares) and jet-powered (2D: blue circles) explosions.}
    \label{fig:tracer_dimen}
\end{figure}

In Figure \ref{fig:tracer_dimen} we plot the tracer thermodynamics for the jet-driven supernova model using the 25 $M_{\odot}$ star as the progenitor, and compare with the spherical explosion model. The difference in the heating history is clear on the tracer particles. (1) In the 1D model, there are large variations in the temperature in the inner core with $\rho_{\rm max} > 10^7$ g cm$^{-3}$. However, there are very small or even no variations for tracers below that density. This is because the shock in the spherical model has lost most of its compressional heating effect at $\rho_{\rm max} < 10^7$ g cm$^{-3}$. (2) On the other hand, the angular effects remain clear to be seen for the 2D model down to $10^5$ g cm$^{-3}$.  This suggests that the energy confined within 15--30$^{\circ}$ helps the shock to maintain its strength to a larger spatial extension. 

\begin{figure}
    \centering
    \includegraphics[width=8.5cm]{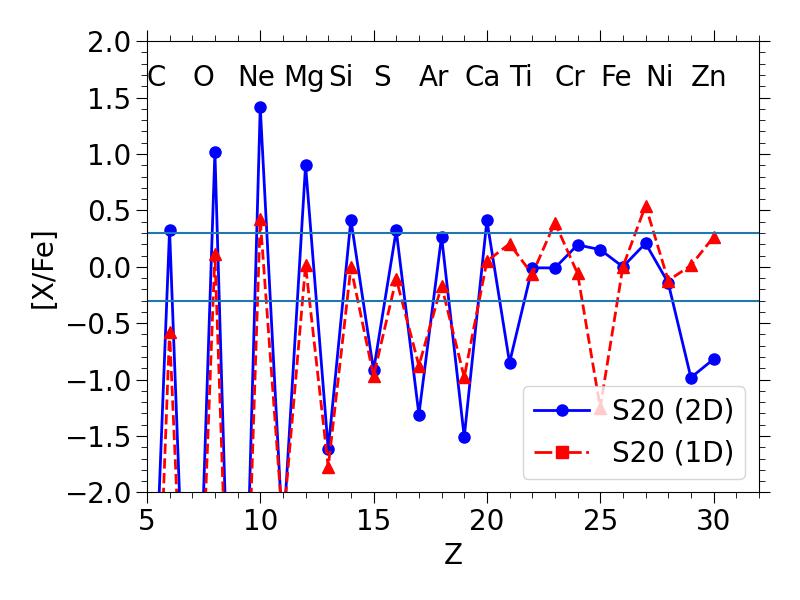}
    \includegraphics[width=8.5cm]{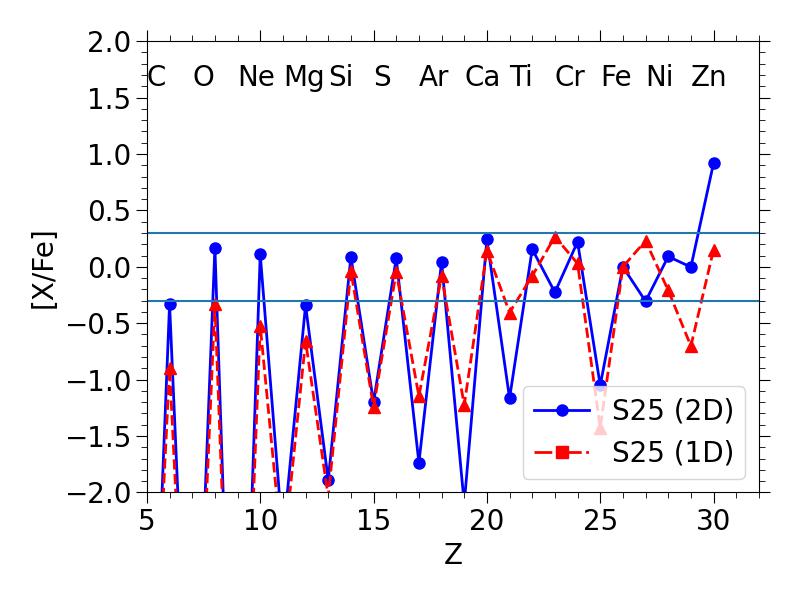}
    \includegraphics[width=8.5cm]{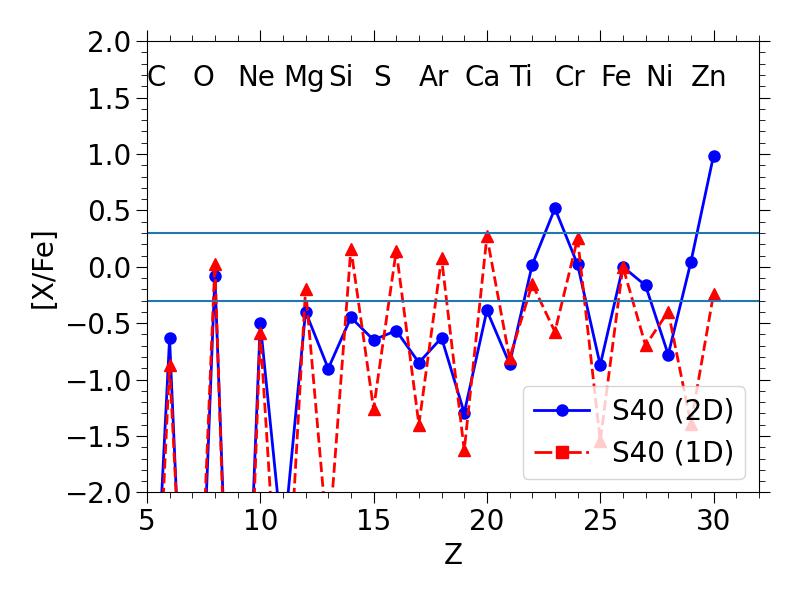}
    \caption{(top panel) The chemical abundances of Models S20-1000-1000-15 (blue solid line) and its one-dimensional counterpart (red dashed line). 
    (middle panel) Same as the top panel but for Models S25-1000-1000-15 and its 1D counterpart.
    (bottom panel) Same as the top panel but for Models S40-1000-1000-15 and its 1D counterpart.
    }
    \label{fig:xiso_dimen_N25}
\end{figure}

In Figure \ref{fig:xiso_dimen_N25} we plot the abundance patterns of our characteristic models using 20, 25 and 40 $M_{\odot}$ progenitors, in comparison with the corresponding spherical (1D) models. Both sequences of models assume the same explosion energy $\sim 1.5 \times 10^{51}$ erg. 

The progenitor mass also affects how the 2D and 1D models differ from each other. For the lower mass model (20 $M_{\odot}$), the 2D explosion leads to the prominent overproduction of O- and Si-group elements along the $\alpha$-chain, $^{16}$O, $^{20}$Ne and $^{24}$Mg. This can be understood by the entire ejection of the outer layers. The odd-number Si-group elements are underproduced in both models, with the exception that Ti and Cr are well-produced in 1D models. The Fe-group elements in both models are similar to the solar ratio, while Ni is significantly overproduced in the 1D model. Zn is also overproduced in the 1D model but is substantially underproduced in the 2D model. 

The 25 $M_{\odot}$ model is like a mixture of the 20 and 40 $M_{\odot}$ models. The O-group elements are weaker for the 1D model while the Si-group elements are stronger. The chemical abundances from Si to Fe are very close to the solar abundances in both dimensionalities, with the exception of the odd-number elements. Ni is again overproduced in the spherical model. The overproduction suggests that, to reconcile with the solar abundance, the ejecta cannot contain the entire inner core of the star. The overproduction is less severe in the 2D model.

The 40 $M_{\odot}$ is in sharp contrast with the $20~M_{\odot}$ model. Most  $\alpha$-chain elements from Ne to Ti in the 2D model are underproduced while the 1D model produces the amounts close to the solar values. The 2D model contains more weights in the Fe-group elements. The ejecta in the 2D model contains a clear Zn signature but a very low Ni abundance, which is an opposite in the 1D counterpart. The even-number charged elements vs. odd-number charged elements (even-odd) parity is also smaller for the 2D model. The mass of even elements are similar in both models but the odd elements are about 10 times higher in the 2D model. This is because in the 2D model, the jet allows the fluid elements which produce Si-group elements ($\rho \sim 3 \times 10^5 - 1 \times 10^7$ g cm$^{-3}$) to have a higher peak temperature when the shock arrives. In our 1D model, the peak temperature reaches about $3.5 \times 10^9$ K; while in the 2D model, it can reach about $4 - 5 \times 10^9$ K. Such a high temperature strongly facilitates the production of odd elements. 

\begin{figure}
    \centering
    \includegraphics[width=8.5cm]{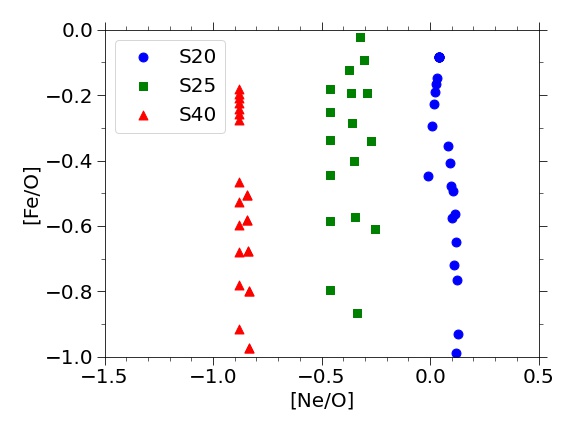}
    \includegraphics[width=8.5cm]{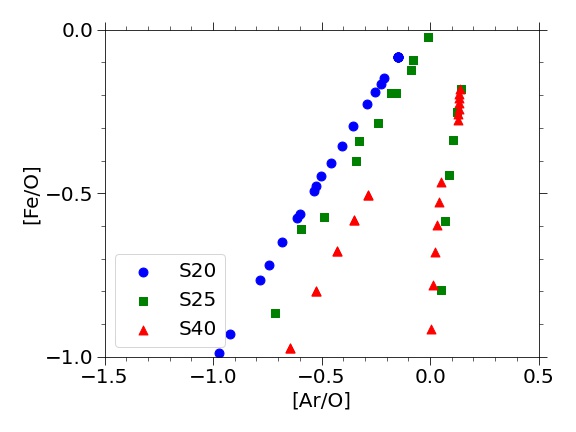}
    \includegraphics[width=8.5cm]{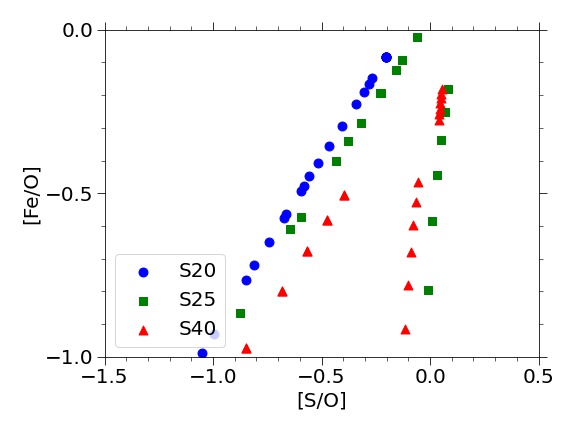}
    \caption{The nucleosynthetic yields of the one-dimensional spherical explosion models with the mixing-fallback mechanism to approximate the aspherical yields. Mixing masses of 3 -- 6 $M_\odot$ are adopted.
    (top panel) [Fe/O] against [Ne/O]. 
    (middle panel) [Fe/O] against [Ar/O]. 
    (bottom panel) [Fe/O] against [S/O].}
    \label{fig:1D_yield}
\end{figure}

\begin{figure}
    \centering
    \includegraphics[width=8.5cm]{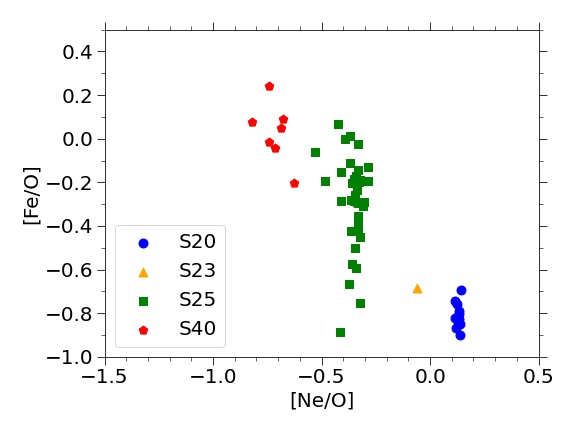}
    \includegraphics[width=8.5cm]{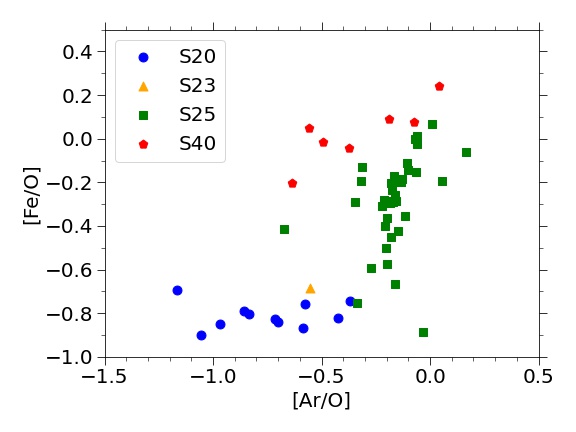}
    \includegraphics[width=8.5cm]{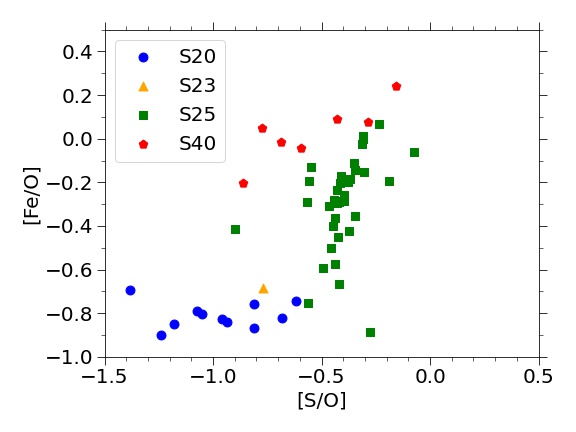}
    \caption{Same as Figure \ref{fig:1D_yield} but for the 2D jet-driven supernova models presented in this work. (top panel) [Fe/O] against [Ne/O]. (middle panel) [Fe/O] against [Ar/O]. (bottom panel) [Fe/O] against [S/O].}
    \label{fig:2D_yield}
\end{figure}

To further visualize the effect of the mixing in one-dimension and the two-dimensional jet-driven explosions, we show in Figure \ref{fig:1D_yield} the one-dimensional spherical model yields using the mixing-fallback mechanism, and in Figure \ref{fig:2D_yield} the two-dimensional model sequences. In all 1D models, we choose the $E_{\rm exp} = 1 \times 10^{52}$ erg as the reference and vary the mixing parameter from $0 - 0.3$ in 0.05 steps. This corresponds to the jet angle $\sim 45^{\circ}$. Different mixing masses are used from 3.0 -- 6.0 $M_\odot$ for each model. We focus on the element pairs [Ne/O], [Ar/O] and [S/O] against [Fe/O] as these elements are frequently observed in galaxies. 

The one-dimensional spherical models in Figure \ref{fig:1D_yield} show vertical variations for different mixing ratios. We note: (1) [Ne/O] in one-dimensional models is a good indicator of the progenitor mass because of the following reason.  Ne is synthesized by C-burning which produces larger $X$(Ne) from larger $X$(C) in the C+O core.  The C/O ratio is smaller for the larger $^{12}{\rm C}(\alpha,\gamma)^{16}$O reaction rate which is higher for higher temperatures.  Since the temperature in the C+O core is higher for the larger progenitor mass, the C/O ratio and thus the Ne/O ratio are smaller for a larger progenitor mass. The ratio does not depend on the mixing level or the mass cut. 

(2) The [Fe/O] of all spherical models are sub-solar because the massive C+O-layer is ejected, which strongly suppresses the ratio. This is in contrast to the 2D model, where a significant fraction of the O-layer is accreted rather than ejected (see Figure \ref{fig:N25_tracers}). The sub-solar patterns persist among all masses. 

(3) For [Ar/O] and [S/O], there are two groups of models. When the mass cut is deep enough to completely isolate the Si-core, these ratios against [Fe/O] appear to be slightly supersolar and vertical (right side of the plot). When the mass cut is moved outward, the ratios appear to align on a straight line. This is because the inner core also contains Ar and S.

The two-dimensional models in Figure \ref{fig:2D_yield} show qualitatively different patterns from their one-dimensional models except [Ne/O].

(1) [Ne/O] against [Fe/O] preserves its strong mass-dependence as in the 1D models.

(2) Other two element pairs show a large scatter. The [Fe/O] ratio is significantly higher than most 1D models. The 25 and 40 $M_{\odot}$ stars have much higher [Fe/O] than most one-dimensional counterpart. 

(3) The 20 $M_{\odot}$ star shows distinctively lower [Fe/O], [Ar/O] and [S/O] than more massive stars, because its more compact progenitor leads to very strong fallback of the Si core as we show in the next section. Such fallback 
suppresses the synthesis and ejection of Fe.  We also observe that [Ar/O] and [S/O] are very similar to each other. The distinction of the one-dimensional approximation and two-dimensional models further indicates the needs of the self-consistent simulations for correctly predicting the mixing and fallback process.


\subsection{Dependence on Progenitor Mass}

\begin{figure}
    \centering
    \includegraphics[width=8.5cm]{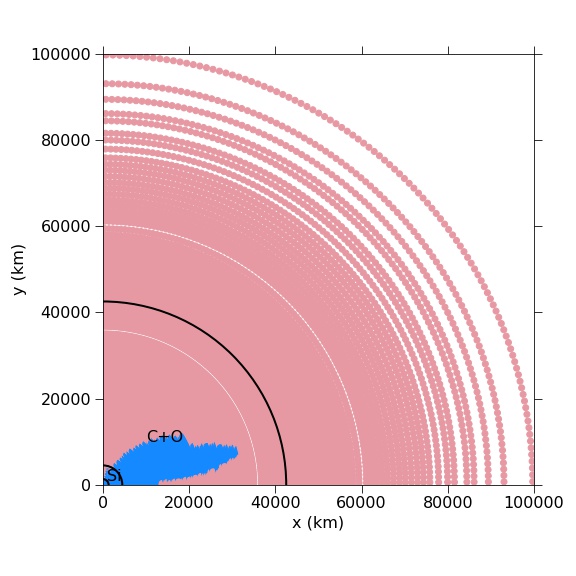}
    \caption{Same as Figure \ref{fig:N25_tracers} but for the characteristic model of 20 $M_{\odot}$ as the progenitor.}
    \label{fig:N20_tracers}
\end{figure}

In Figure \ref{fig:N20_tracers} we present the distribution of the tracer particles for the 20 $M_{\odot}$ model using the standard jet energetics. The progenitor has thin Si and C+O layers at the onset of collapse. Despite this model has a lower binding energy than the 25 $M_{\odot}$ model counterpart, a larger fraction of the Si layer remains bound at the end of simulation. We note that both models receive the same amount of the deposited energy. The inner C+O layer is ejected in an aspherical manner. The cone shaped ejecta, with the energy deposition being confined within $15^{\circ}$, can extend beyond $45^{\circ}$. Matter above $\sim$ 15,000 km is fully ejected.

\begin{figure}
    \centering
    \includegraphics[width=8.5cm]{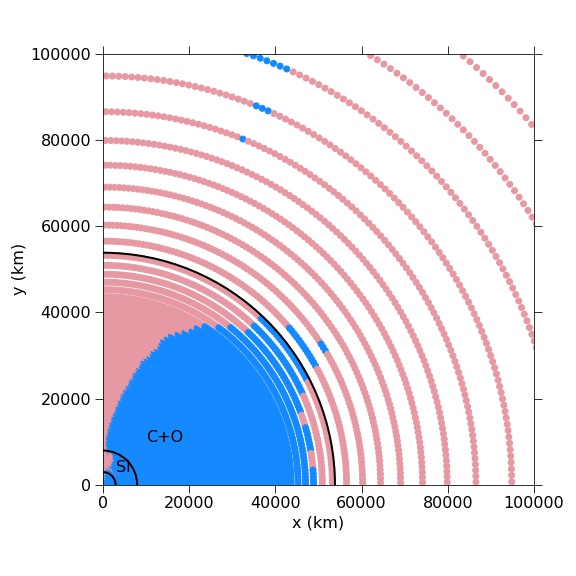}
    \caption{Similar to Figure \ref{fig:N25_tracers} but for the characteristic model of 40 $M_{\odot}$ as the progenitor.}
    \label{fig:N40_tracers}
\end{figure}

Similar to the above plot, we plot the tracers for the 40 $M_{\odot}$ model in Figure \ref{fig:N40_tracers}. The tracers which can escape show a clearer jet structure. It can be separated into two parts. Below $r \sim 30,000$ km, the tracers are limited to the cone similar to the jet, but with a wider open angle $\sim 30^{\circ}$. Above 30,000 km, ejecta appears to be spherical in distribution, with the exception near the boundary.

\begin{figure}
    \centering
    \includegraphics[width=8.5cm]{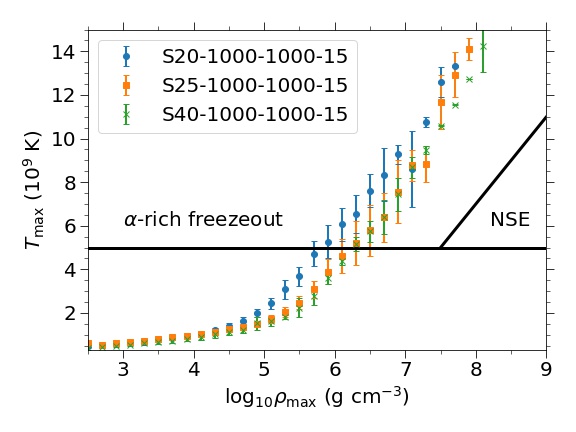}
    \caption{Same as Figure \ref{fig:tracer_benchmark_S25} but for the 20 $M_{\odot}$ and 40 $M_{\odot}$ progenitor models with the standard jet. }
    \label{fig:tracer_stat_mass}
\end{figure}

In Figure \ref{fig:tracer_stat_mass} we show a plot of tracers being similar to Figure \ref{fig:tracer_benchmark_S25} but including models with different progenitor masses. All three models show monotonically increasing distributions of $T_{\rm max}$ against $\rho_{\rm max}$, suggesting that the shock is propagating only outward radially, without observable reflection or collision. The 25 and 40 $M_{\odot}$ models are less compact than the 20 $M_{\odot}$ model but show similar tracer distributions.
The temperature fluctuations at $\rho = 10^5 - 10^7$ g cm$^{-3}$ are smaller for 25 and 40 $M_{\odot}$ models than 20 $M_{\odot}$.  All models show that at $\rho > 10^6$ g cm$^{-3}$ the tracers experience only $\alpha$-rich freezeout.

\begin{figure}
    \centering
    \includegraphics[width=8.5cm]{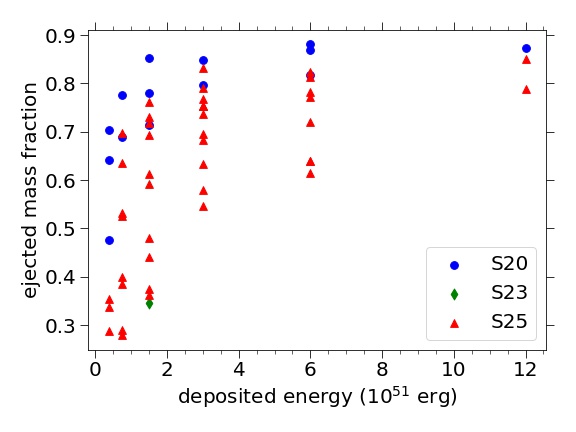}
    \caption{The ejecta mass fraction of the models in Tables \ref{table:S20models} and \ref{table:S25models}}
    \label{fig:ejecta_frac}
\end{figure}

In Figure \ref{fig:ejecta_frac}, we plot the mass fraction of the ejecta. The ejecta mass fraction is the mass fraction of passive tracer particles which have a positive total energy by the end of the simulation. These tracers, which are also post-processed for nucleosynthesis, are the matter to be ejected in the explosion. For each progenitor mass, the ejecta fraction tends to increase as the deposited energy
increases. There are large variations among models with the same deposited energy because the robustness of the ejection depends the shock strength, which consequently depends on the jet properties.

The ejecta mass fraction of the S20-series reaches the asymptotic value of $\sim$ 0.85 at a relatively low deposited energy of $\sim 2 \times 10^{51}$ erg. On the other hand, the more massive progenitor of S25-series requires a higher deposited energy of $\sim 3 \times 10^{51}$ erg for the ejecta mass fraction to reach the asymptotic value of $\sim$ 0.80. The lower asymptotic value can be understood because the more massive Si-core tends to be accreted regardless of the deposited energy. We remind that due to the jet geometry, the equatorial matter in the Si-core is
less likely to be ejected.


\begin{figure}
    \centering
    \includegraphics[width=8.5cm]{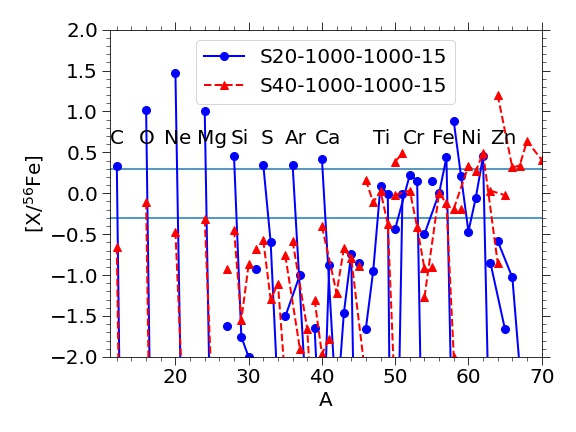}
    \caption{Chemical abundance patterns of Models S20-1000-1000-15 (20 $M_\odot$: blue solid line) and S40-1000-1000-15 (40 $M_\odot$: red dashed line).}
    \label{fig:N25_mass_xiso}
\end{figure}

In Figure \ref{fig:N25_mass_xiso} we plot the isotopic abundances of two contrasting models S20-1000-1000-15 (20 $M_\odot$) and S40-1000-1000-15 (40 $M_\odot$). The two models share the same jet energy but different progenitor masses. The differences of the pre-collapse structure in terms of the relative mass fractions of the C+O and Si-layers, together with their binding energies, lead to very substantial differences in the abundance patterns. 

The lower mass progenitor (20 $M_\odot$) produces such large ratios of O-group and Si-group isotopes relative to $~{56}$Fe as [($^{16}$O,$^{20}$Ne,$^{24}$Mg)/$^{56}$Fe] $\gtrsim$ 1.0, [($^{28}$Si,$^{32}$S,$^{36}$Ar,$^{40}$Ca)/$^{56}$Fe] $\gtrsim$ 0.3 (and thus [($^{16}$O/$^{24}$Si] $\gtrsim$ 0.2).

In contrast, the higher mass progenitor (40 $M_\odot$) produces much smaller ratios of O-group and Si-group isotopes relative to $~{56}$Fe, i.e., [($^{16}$O,$^{20}$Ne,$^{24}$Mg)/$^{56}$Fe] $\lesssim -0.1$,
[($^{28}$Si,$^{32}$S,$^{36}$Ar,$^{40}$Ca)/$^{56}$Fe] $\lesssim -0.4$ (and thus [($^{16}$O/$^{24}$Si] $\gtrsim$ 0.3).


For Fe-group isotopes, the differences are irregular. For example, the lower mass model shows a suppressed $^{64}$Zn production but an enhanced $^{58}$Ni production relative to $^{56}$Fe. In contrast, the higher mass model shows opposite ratios.  Such a difference is due to the difference in the $Y_e$ distribution.



\subsection{Effects of Mass Cut}

\begin{figure}
    \centering
    \includegraphics[width=8.5cm]{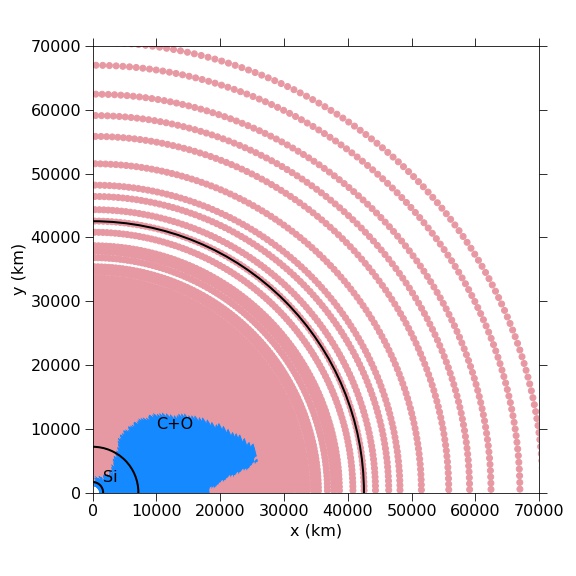}
    \caption{Similar to Figure \ref{fig:N20_tracers} but for the mass cut at $R_{\rm cut} =$ 1500 km in the 25 $M_{\odot}$ model with the same jet parameters.}
    \label{fig:N25_Rcut1500_tracers}
\end{figure}

\begin{figure}
    \centering
    \includegraphics[width=8.5cm]{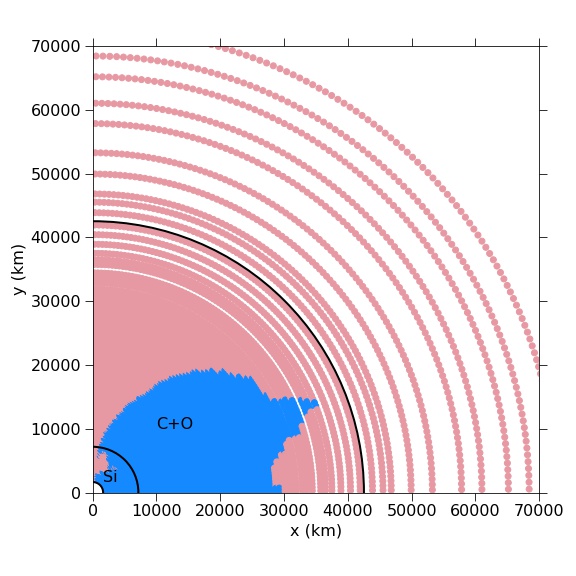}
    \caption{Similar to Figure \ref{fig:N20_tracers} but for the mass cut at $R_{\rm cut} =$ 2100 km in the 25 $M_{\odot}$ model with the same jet parameters.}
    \label{fig:N25_Rcut2100_tracers}
\end{figure}

In Figures \ref{fig:N25_Rcut1500_tracers} and \ref{fig:N25_Rcut2100_tracers}, we plot the tracer particle distribution for the 25 $M_{\odot}$ star being similar to Figure \ref{fig:N25_tracers} but with the different mass cut (i.e., inner boundary) at radii of $R_{\rm cut} =$ 1500 km
and 2100 km, respectively. A direct comparison with Figure \ref{fig:N25_tracers} shows a drastic difference in the ejecta mass and structure. While the two models have the same jet energy, the 
mass cut at the larger $R_{\rm cut}$ leads to more significant fallback including both the Si-core and the inner C+O-core. The ejection of the He-envelope is only mildly reduced.


The change of the mass cut to larger $R_{\rm cut}$ causes the following changes in the ejecta structure. 

First, the mass cut at larger $R_{\rm cut}$ means the formation of a more massive compact object. 
Second, the larger $R_{\rm cut}$ leads to more mass receiving the jet energy. The final energy per unit mass after the jet deposition is lower, which means a weaker outgoing shock. The effect of $R_{\rm cut} =$ 2100 km is significant. Even with the same amount of deposited energy, the majority of the inner core fails to be ejected. Only the outskirt of the C+O core is ejected at the end of simulation. 

\begin{figure}
    \centering
    \includegraphics[width=8.5cm]{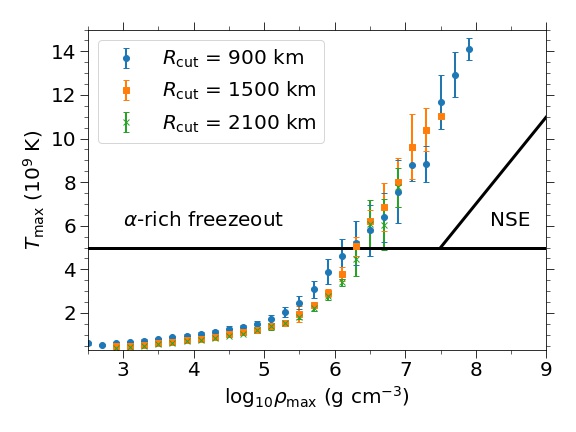}
    \caption{The tracer thermodynamics history of the characteristic model using the 25 $M_{\odot}$ star as the progenitor with $R_{\rm cut} = $ 900, 1500 and 2100 km at the mass cut.}
    \label{fig:N25_Rcut_thermo}
\end{figure}

In Figure \ref{fig:N25_Rcut_thermo} we show the thermodynamics history of our characteristic models using the 25 $M_{\odot}$ progenitor, but with the innermost radius of 900, 1500 and 2100 km, respectively. The maximum density experienced by the tracers is smaller for the larger inner radius. For the model with $R_{\rm cut} = $ 2100 km, all tracers have densities less than $10^7$ g cm$^{-3}$. 

\begin{figure}
    \centering
    \includegraphics[width=8.5cm]{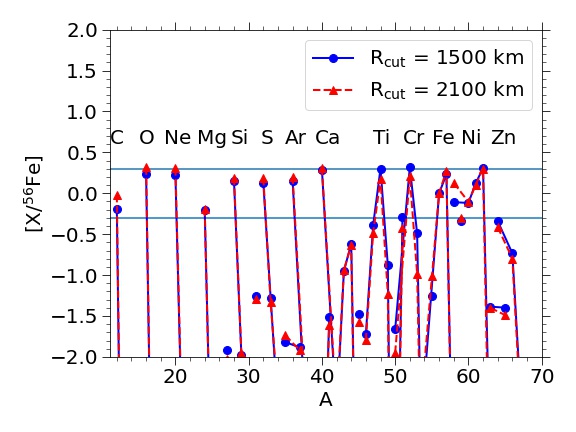}
    \caption{The chemical abundance patterns of Model S25-1000-1000 with $R_{\rm cut} = 1500$ km (blue solid line) and $R_{\rm cut} = 2100$ km (red dashed line).}
    \label{fig:N25_Rcut_xiso}
\end{figure}

In Figure \ref{fig:N25_Rcut_xiso}, we compare the abundance patterns with different mass cuts. It does not show qualitative differences. The two models follow each other, with the model with a more extended mass cut (1500 km) have in general 20 -- 50\% higher in O-group and Si-group elements up to Ca. The relation flips for more massive elements. One exception is $^{50}$Cr and $^{54}$Cr. The more extended mass cut can boost its production. The mild difference of the Fe-group elements, mostly from the innermost region of the star, is small in terms of the mass cut. This agrees with the ejecta distribution that the narrow segment in the Si-core is ejected in both models. Also, the accreted matter is mostly C+O-rich matter which does not bring major nucleosynthetic changes to other elemental abundances.

\section{Discussion}
\label{sec:discussion}

\subsection{The Ti-V relation Revisited}

\begin{figure}
    \centering
    \includegraphics[width=8.5cm]{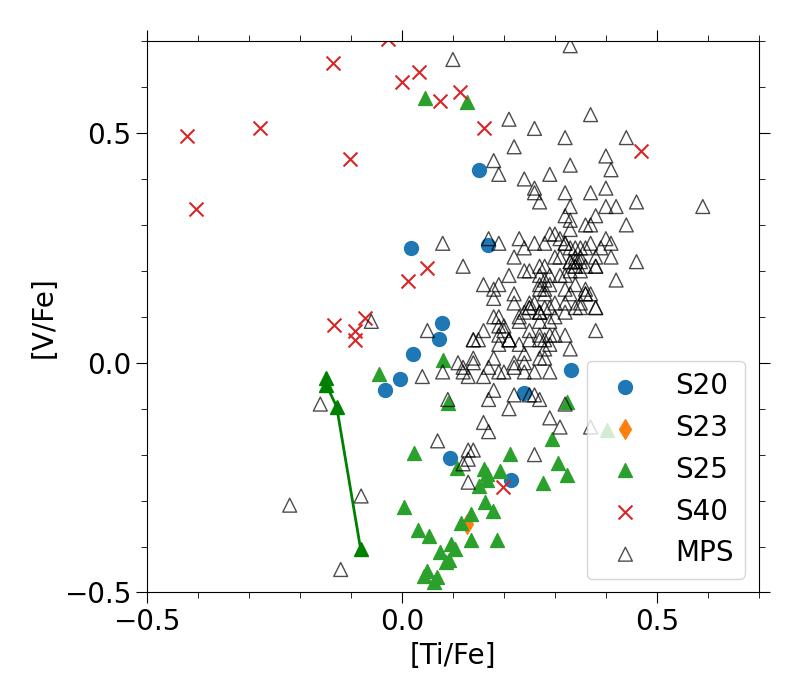}
    \caption{The [V/Fe] against [Ti/Fe] relation for the jet-driven supernova models in this work. The semi-transparent purple diamonds stand for the metal-poor star data taken from \citep{Roederer2014}. The straight line connects the spherical explosion models of S25 with different explosion energies.}
    \label{fig:Ti_V_plot}
\end{figure}

\cite{Sneden2016} have found that [Ti/Fe] and [V/Fe] show a correlation of 45$^{\circ}$ based on the metal-poor stars catalogue derived in \cite{Roederer2014} (Fig. \ref{fig:Ti_V_plot}). In \cite{Leung2023Jet1}, we have explored the [Ti/Fe] $-$ [V/Fe] relation using the 40 $M_{\odot}$ models and compared with the observed relation. 
In our previous work, the 45$^{\circ}$ relation is reproduced but the interception differs from the data.

In Figure \ref{fig:Ti_V_plot}, we extend our comparison with the available data by including our new models.
Being consistent with our 40 $M_{\odot}$ models, there is also a clear trend in our 25 $M_{\odot}$ models which also show a 45$^{\circ}$ inclination in the [Ti/Fe]--[V/Fe] plane. The 20 $M_{\odot}$ series of models, on the contrary, does not show a clear trend but their composition is closer to the cluster formed by the observational data. The 40 $M_{\odot}$ models are in general very remote to those EMP stars. They have a much higher [V/Fe] ratio. 

To compare with the spherical explosion models, we also show one particular mass of $25~M_{\odot}$ with different explosion energy. It shows that in the high energy limit, the spherical model resembles with our aspherical models. This agrees with our expectation that the jet model can create high-entropy matter which is similar to a very strong explosion in the spherical case. 

If we extend the data assuming a linear trend, the intercepts of models with different progenitor mass does not vary monotonically. It is because in the lower mass models, the fallback becomes important that the Fe production is also affected. It suggests a non-linear competition between lowering the mass and the enhanced production of [Ti/Fe]. A further inspection will be useful for exploring the mass dependence of this element pair. 

\subsection{$^{56}$Ni mass and ejecta mass relation}

An interesting relation for Type Ib/c supernovae is the correlation between the ejected $^{56}$Ni mass ($M_{56}$) against the ejecta mass ($M_{\rm ej}$). These two quantities can be well extracted from the light curve. Qualitatively, the peak brightness and the width of the light curve determines the $M_{56}$ and $M_{\rm ej}$, respectively. In the iPTF (Intermediate Palomar Transient Factory) surveys presented in \cite{DeCia2018, Gomez2022, Taddia2019}, a clustering is observed for this kind of supernovae. Here we examine if a similar relation can be reproduced for the jet-driven models. 

\begin{figure}
    \centering
    \includegraphics[width=8.5cm]{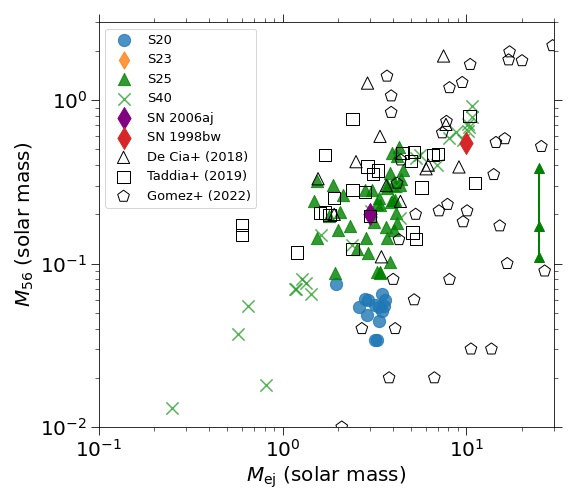}
    \caption{The ejected $^{56}$Ni mass ($M_{56}$) against the ejecta mass ($M_{\rm ej}$) of the jet-driven supernova models in this work (20 $M_{\odot}$ series: blue circles, 23 $M_{\odot}$-series: orange diamonds, 25 $M_{\odot}$-series: green triangles, and 40 $M_{\odot}$-series: green crosses). The plot is overlaid with the observed Type Ib/c supernova catalog taken from \cite{DeCia2018} (brown triangles), \cite{Taddia2019} (pink squares), \cite{Gomez2022} (grey pentagons). The green triangles connected with the line are the spherical explosions of the S25 models with different explosion energies.}
    \label{fig:mni_mej_plot}
\end{figure}

In Figure \ref{fig:mni_mej_plot}, we plot $M_{56}$--$M_{\rm ej}$ relation for the catalog in this work and the surveys described above. It shows that models with different progenitor masses have significant overlap in the parameter space as follows. (1) The S20 series does not exhibit large variations and it is suitable to explain the faint supernovae of $M_{56} < 0.1~M_{\odot}$ with $M_{\rm ej} = $ 0.1 -- 0.5 $M_{\odot}$. (2) The S40 series has a clear rising trend of $M_{56}$. (3) The S25 series has $M_{56}$ less sensitive to $M_{\rm ej}$. The contrast among these series is related to the compactness of the progenitor model. In a lower mass model, the higher compactness leads to strong fallback in the inner core. 

By observing how the theoretical data overlap with the observational data, we conclude that the 25 $M_{\odot}$ series agrees with the observed Type Ib/c supernovae from \cite{Taddia2019} and \cite{Gomez2022}. These supernovae (including the well observed SN 1998bw and 2006aj) have typically $M_{56} \sim 0.1 - 1~M_{\odot}$ and $M_{\rm ej} = 1 -10~M_{\odot}$. However, the jet-driven supernova cannot explain directly the data point with a $M_{\rm ej} >0.5~M_{\odot}$) and a low $M_{56}$ ($<0.1~M_{\odot}$). Similarly, this model cannot fully explain supernovae with high $M_{56}$ ($>1~M_{\odot}$).

We also include the spherical explosion model for the S25 progenitor shown the triangles with a line. The high explosion energy makes the entire envelope ejected. Thus, $M_{\rm ej}$ does not change significantly across different explosion energy. $M_{56}$ increases sharply as the explosion energy increases. The high $M_{\rm ej}$ does not well resemble to the clustering of the data. 

\subsection{Elemental Abundance of Jet-Driven Supernovae}

In the previous sections, we have presented the detailed isotopic abundances of our models. In this section, we compare our results with some well-observed metal-poor galaxies. These galaxies, some of these having relatively low metallicity, are good test cases for our zero metallicity models because such early galaxies might have a much lower fraction of metal enrichment by Type Ia supernovae. Most explosions are contributed by either spherical and aspherical massive star explosions. Therefore, it reduces the dependence on some of the uncertain parameters in Type Ia supernovae, such as the explosion channel \citep{Leung2021SNIaReview} and the delay time \citep{Kobayashi2020}.  

\begin{figure}
    \centering
    \includegraphics[width=8.5cm]{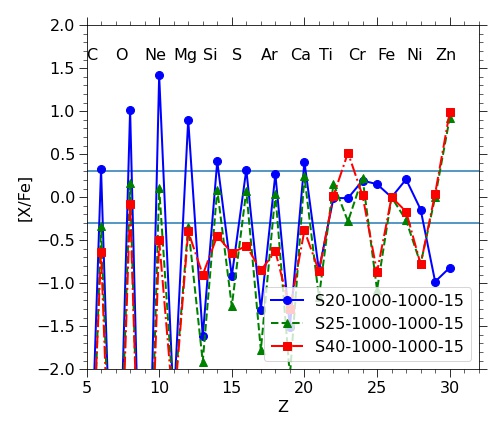}
    \caption{The elemental abundances [X/Fe] of our characteristic models using 20 $M_{\odot}$ (blue solid line, circle), 25 $M_{\odot}$ (green dashed line, triangle) and 40 $M_{\odot}$ (red dot-dashed line, square) models. The two lines correspond to 50\% and 200\% of the solar value.}
    \label{fig:jet_mass_elem}
\end{figure}

In Figure \ref{fig:jet_mass_elem} we aggregate the abundances element-wise and plot the elemental abundance patterns. The graph provides a direct comparison with observational data because the spectral lines from galaxies do not distinguish the detailed distribution across stable isotopes. Nonetheless, we stress that the isotopic distribution is equally important because it allows us to clarify the explosion physics quantitatively. 

Figure \ref{fig:jet_mass_elem} shows that [(C-Ca)/Fe] $> 0.3$ in the 20 $M_{\odot}$ model, while [(C-Ca)/Fe] $< -0.1$ in the 40 $M_{\odot}$ model. The 25 $M_{\odot}$ model has the chemical abundance most proximate to the solar abundance pattern. All three models show a comparable Fe-group element production, with the exception near Zn. The higher mass model tends to produce more Zn than the lower mass models. 

\section{Comparison with Observational Data}
\label{sec:mps}

\subsection{Extremely Metal Poor Galaxies}

\begin{table*}[]
    \caption{The chemical abundance data of some extremely metal-poor galaxies collected from recent works as reported in \cite{Watanabe2023}.}
    \centering
    \begin{tabular}{c c c c c c}
        \hline
         Galaxy ID & [Ne/O] & [S/O] & [Ar/O] & [Fe/O] & Reference \\
         \hline
         SBS-0335-052E & $-0.27 \pm 0.00$ & $0.018 ^{+0.05}_{-0.018}$ & $-0.037 ^{+0.02}_{-0.03}$ & $-0.24^{+0.11}_{-0.24}$ & \cite{Watanabe2023} \\
         J0125+0759 & $-0.01 \pm 0.00$ & $-0.34 \pm 0.05$ & $-0.08 ^{+0.01}_{-0.08}$ & $-0.24^{+0.12}_{-0.24}$ & \cite{Watanabe2023} \\
         J0159-0622 & $-0.12 ^{+0.01}_{-0.02}$ & $-0.016 \pm 0.06$ & $-0.01 \pm 0.02$ & $-0.51^{+0.17}_{-0.28}$ & \cite{Isobe2022} \\
         J1608+4337 & $0.107 \pm 0.02$ & $-0.07 \pm 0.02$ & $-0.04 ^{+0.02}_{-0.03}$ & $-0.43^{+0.15}_{-0.22}$ & \cite{Isobe2022}\\
         J2115-1734 & $0.003 ^{+0.004}_{-0.005}$ & $-0.079 \pm 0.06$ & $0.04 ^{+0.01}_{-0.06}$ & $-0.41^{+0.03}_{-0.02}$ & \cite{Kojima2020} \\
         J0811+4730 & $-0.0048 \pm 0.031$ & $-0.117 \pm 0.056$ & $-0.245 \pm 0.104$ & $0.17 \pm 0.092$ & \cite{Izotov2018} \\ \hline
    \end{tabular}
    
    \label{table:watanabe_EMPG}
\end{table*}

In \cite{Watanabe2023} the EMPRESS survey continues to analyze the chemical abundances of some metal-poor galaxies. The analysis includes two new objects SBS-0335-052E and J0125+0759. In Table \ref{table:watanabe_EMPG} we list the abundance data of these galaxies, together with some other metal-poor galaxies collected from relevant surveys. Similar to \cite{Mao2021}, Ne, Ar, S and Fe relative to O are measured. We list the galaxies we have used for our comparison in Table \ref{table:watanabe_EMPG}.

Table \ref{table:Isobe_EMPG} lists the EMPG studied in this work. We focus on the samples which have good measurements of the Ne/O and Ar/O ratios. Among these EMPGs, they share similar values of $\log_{10}$ Ar/O $\sim -2.3$, which is significantly sub-solar. Their Ne/O ratio $\log_{10}$ Ne/O $\sim -0.6$ is slightly sub-solar. The relatively flat distribution of the Ar/O ratio among these galaxies suggest that these galaxies have experienced similar metal-enrichment history. 

\begin{table}[]
    \caption{Elemental abundance data of EMPGs used in this work, as extracted in \cite{Isobe2022}.}
    \centering
    \begin{tabular}{c c c c}
    \hline 
         EMPG &  $\log_{10}$(Ne/O) & $\log_{10}$(Ar/O) & $\log_{10}$(Fe/O) \\ \hline 
         J0156-0421 & -0.759 & -2.39 & $<-0.45$ \\
         J0159-0622 & -0.878 & -2.32 & -1.74 \\
         J0210-0124 & -0.688 & -2.37 & -1.75 \\
         J0226-0517 & -0.664 & -2.38 & -1.27 \\
         J0232-0248 & -0.692 & -2.42 & -1.26 \\ \hline 
    \end{tabular}
    \label{table:Isobe_EMPG}
\end{table}

We should note that, for abundance comparisons between EMPGs and theoreical models, we need chemical evolution models of galaxies as mentioned in \cite{Watanabe2023}.  In the present work, we make a simple comparison between the $\sim 20-40~M_{\odot}$ Pop III stars and EMPGs in order to study how important the jet-like explosions are for chemically enriching EMPGs.  Also, $\sim 20-25~M_{\odot}$ stars have
been shown to make the largest contribution in chemical evolution of galaxies \citep{Weaver1978, Woosley1995}.

\begin{figure}
    \centering
    \includegraphics[width=8.5cm]{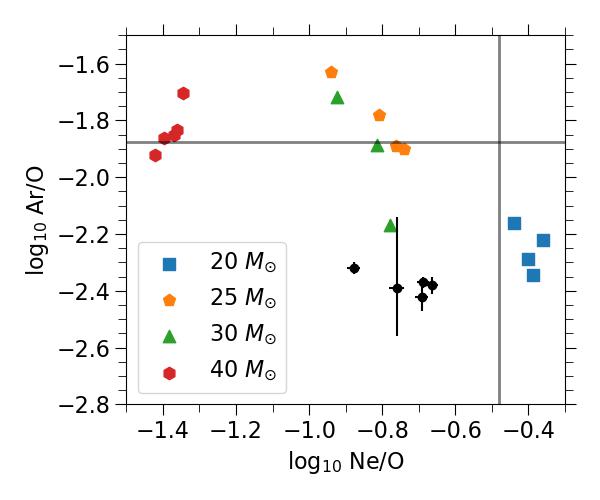}
    \caption{The comparison of the post-explosion [Ne/O] against [Ar/O] for selected EMPGs and the ``spherical'' explosion models.  
    The straight line stands for the solar values of the corresponding isotope ratios.
    The data points correspond to the EMPGs reported in \cite{Isobe2022}.}
    \label{fig:MPG_spherical}
\end{figure}

In Figure \ref{fig:MPG_spherical} we compare the post-explosion nucleosynthesis of our spherical models from 20 -- 40 $M_{\odot}$ for different explosion energies. We demonstrate the mismatch between the ``spherical'' explosion models and EMPGs. We note that the smaller mass star tends to produce larger [Ne/O] with a weak dependence on the explosion energy. And the higher energy explosion produces smaller [Ar/O].

The progenitor mass dependence of [Ne/O] stems from the C/O ratio after He burning which is smaller for a larger mass because of the higher temperature during He burning. The trend of [Ar/O] is expected because a higher explosion energy implies a stronger shock and a higher post-shock temperature, which burn more O into Si, S, and Ar. This facilitates the fusion of the C+O core into elements beyond Ar. Among all models, only the models $\sim 20-30~M_{\odot}$ have [Ne/O] close enough to the observational data. However, their [Ar/O] is too high with the exception of one $30_{\odot}$ model of low explosion energy.

\begin{figure}
    \centering
    \includegraphics[width=8.5cm]{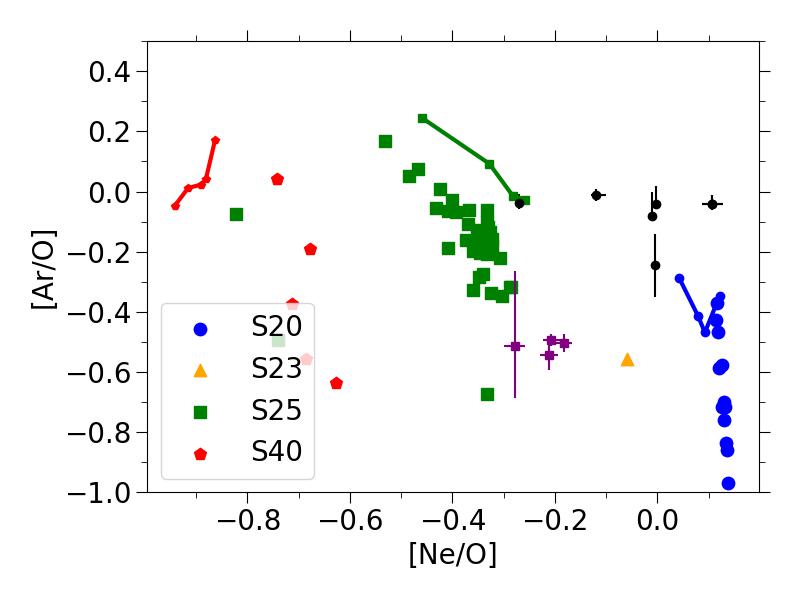}
    \caption{[Ar/O] against [Ne/O] for the jet-driven supernova models in this work, compared with the metal poor galaxy (EMPG) catalog presented in \cite{Watanabe2023} (black circles with error bars) and in \cite{Isobe2022} (purple squares with error bars). The coloured lines stand for the sequence of 1D models with the different explosion energies.}
    \label{fig:watanabe_ArO}
\end{figure}

In Figure \ref{fig:watanabe_ArO}, [Ar/O] against [Ne/O] is plotted for the jet-driven models presented in this work and the galaxy (EMPG) data. The horizontal spread of the observed data shows that [Ar/O] is insensitive to [Ne/O], and is higher than the theoretical models. It implies that there are some input physics in the stellar evolutionary models which  leads to a systematically smaller [Ar/O]. [Ne/O] has a very small uncertainty. Based on their values, we can infer that the peak progenitor mass that contributes to the observed abundance is between 20 and 25 $M_{\odot}$. 
[Ne/O] of these EMPGs suggest that they have the supernova contribution centered around 23 $M_{\odot}$ stars in their chemical evolution. 

\begin{figure}
    \centering
    \includegraphics[width=8.5cm]{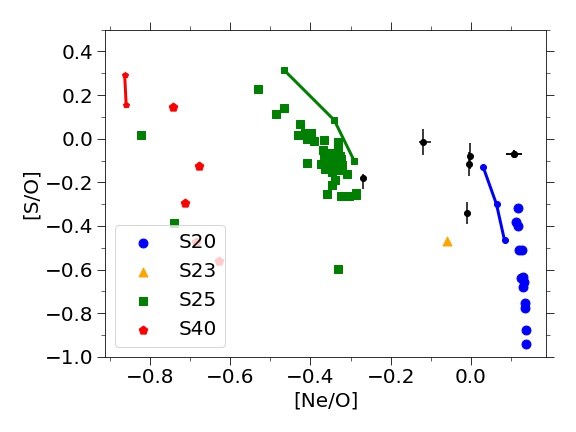}
    \caption{Same as Figure \ref{fig:watanabe_ArO} but for [S/O] against [Ne/O].}
    \label{fig:watanabe_SO}
\end{figure}

\begin{figure}
    \centering
    \includegraphics[width=8.5cm]{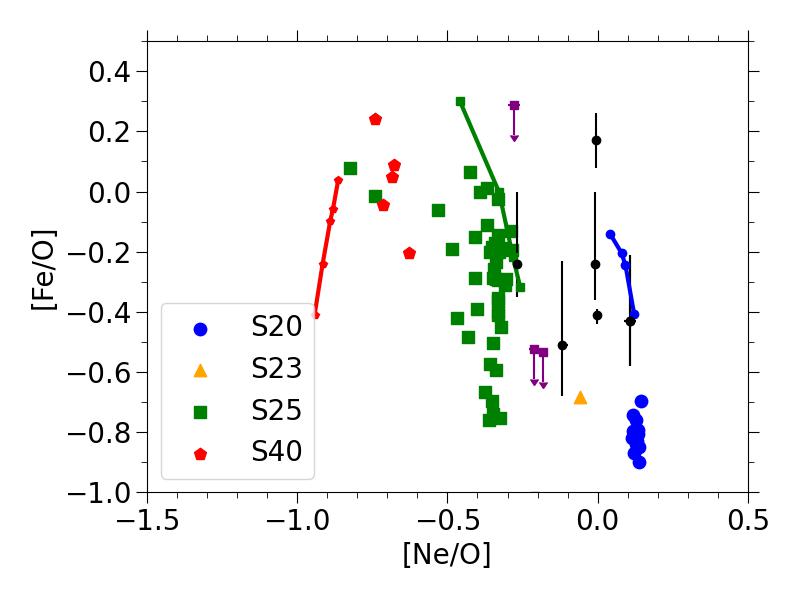}
    \includegraphics[width=8.5cm]{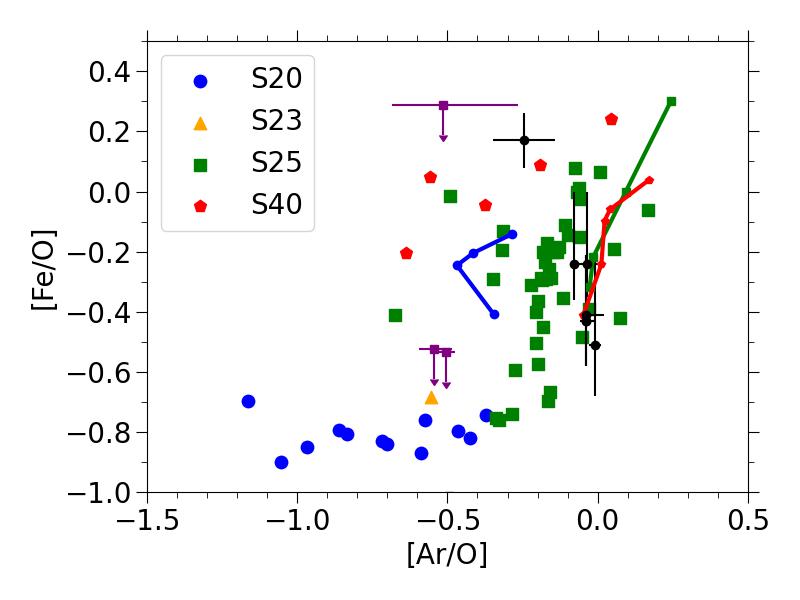}
    \includegraphics[width=8.5cm]{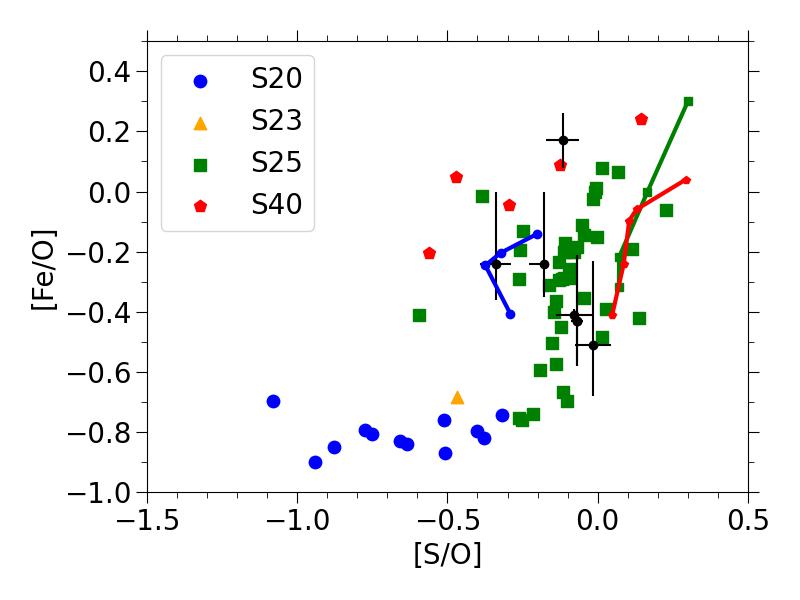}
    \caption{(top panel) Same as Figure \ref{fig:watanabe_ArO} but for [Fe/O] against [Ne/O]. (middle panel) Same as the top panel but for [Fe/O] against [Ar/O]. (bottom panel) Same as the top panel but for [Fe/O] against [S/O].}
    \label{fig:watanabe_FeO}
\end{figure}

We also compare our models with the data points from \cite{Isobe2022} (purple squares with error bars). This element pair also demonstrates a clear clustering for this dataset. The clustering of the data point suggests that these galaxies have a large contribution of the $25~M_{\odot}$ star explosion in their chemical evolution. The comparison can be similarly made for the data points from \citep{Watanabe2023} (black circles with error bars). These data points have [Ne/O]$\sim -0.1$ -- 0.1, which lie somewhere between the clusters of 20 and 25 $M_{\odot}$ star explosions. Their high [Ar/O] suggests that these EMPGs require contributions from the high energy explosions of these two groups. 

In Figure \ref{fig:watanabe_SO}, we plot [S/O] against [Ne/O] for the same sequences of input models. The observational data show that both ratios have a narrow range of $-0.4$ -- 0.0, being similar to the solar abundance. The range of [S/O] is consistent with the jet-driven models. The spread of the S25 models is similar to that for the observed galaxies. 

In the top panel of Figure \ref{fig:watanabe_FeO}, we plot [Fe/O] against [Ne/O] for the same sequences of input models. The observed [Fe/O] shows larger uncertainties compared to the [Ar/O] and [S/O] data. It is interesting to note that, despite the extremely low metallicity in these galaxies, their [Fe/O] does not significantly differ from the solar value. The jet-driven models again produce a spread consistent with the observed range of the abundances ratios. 


In the middle panel of the same figure, we compare [Fe/O] against [Ar/O], alongside with the observational data. Different from [Ne/O], the numerical data cluster at [Ar/O] $= -1.0$ to 0. A smaller mass model tends to have a lower  [Ar/O]. The numerical models coincide well with galaxy abundances from \cite{Watanabe2023}. There are three galaxies reported in \cite{Isobe2022} and all of them show [Ar/O] around -0.5. The data with low [Fe/O] overlap with the 20 $M_{\odot}$ spherical explosion models.  

The figure can be further compared with the first panel of Figure 2 of \cite{Watanabe2023}. In their work, they compare various massive star explosion models with the galaxies they observed. They show that the spherical models requires parametrized ``mixing-and-fallback'' in order to reproduce the observed abundance ratio, or require contributions of Type Ia supernovae to match the observed [Fe/O]. Our models show that with the jet-driven explosion, the theoretical models may reach the observed high value of [Fe/O].

In the third panel of the same figure, we plot [Fe/O] against [S/O] as in the second panel of Figure 2 of \cite{Watanabe2023}.  The numerical models show a scatter of [S/O] being similar to [Ar/O], because both elements are Si-group elements. We notice that a lower mass model generates a smaller [S/O]. The 1D models tend to synthesize too much [S/O] but not enough [Fe/O] in this parameter space. In contrast, some galaxy samples match very well with the jet-driven explosion models. 

By cross-examining these three figures, it suggests that the aspherical explosion could provide the clue to understand the unique chemical abundance patterns of these metal-poor galaxies.

\subsection{Extremely Metal Poor Stars}

Another ideal test case for massive star nucleosynthesis is extremely metal-poor stars (EMPS). Similar to extremely metal-poor galaxies, the elemental abundances of these old stars correspond to one or a few massive star explosions. The contribution of Type Ia supernovae remains less important. However, due to their small length scale, the inhomogeneous mixing and the angle dependence of the aspherical explosion could be important.

Recently, \cite{Jeong2023} made analysis of the high-resolution spectra from stars by the legacy Sloan Digital Sky Survey \citep[SDSS][]{York2000} and the Large sky
Area Multi-Object Fiber Spectroscopic Telescope \cite[LAMOST][]{Cui2012}. They identified 18 very metal-poor stars (VMPS), 10 EMPS and 3 carbon-enhanced metal-poor stars (CEMPS), and perform follow-up observations with the Gemini Remote Access to the CFHT ESPaDOnS Spectrograph \citep[GRACES, ][]{Chene2021}. We selected the stars with complete measurement of Mg, Ca, Cr and Ni in Table \ref{table:jeong_stars} with a list of stars taken from their catalogue.

\begin{table*}[]
    \caption{Selected extreme metal-poor stars from \cite{Jeong2023}. The classification includes VMPS (very metal-poor stars: [Fe/H]$<-2$), EMPS (extremely metal-poor stars: [Fe/H]$<-3$), and CEMPS (carbon-enhanced metal-poor stars: [C/Fe]$>1.0$).}
    \centering
    \begin{tabular}{c c c c c c c}
    \hline
     EMP ID & [Fe/H] &  [Mg/Fe] & [Ca/Fe] & [Cr/Fe] & [Ni/Fe] & Classification \\ \hline
     J0010 & $-2.48\pm0.11$ & $0.73\pm0.04$ & $0.19\pm0.05$ & $-0.10\pm0.05$ & $0.13\pm0.05$ & VMPS \\
     J0158 & $-3.04\pm0.05$ & $0.06\pm0.06$ & $0.41\pm0.04$ & $-0.09\pm0.06$ & $0.30\pm0.06$ & EMPS \\
     J0357 & $-2.75\pm0.05$ & $0.34\pm0.04$ & $0.33\pm0.03$ & $-0.16\pm0.05$ & $0.37\pm0.07$ & EMPS \\
     J0713 & $-3.15\pm0.08$ & $0.31\pm0.06$ & $0.00\pm0.05$ & $-0.33\pm0.03$ & $-0.01\pm0.04$ & EMPS \\
     J0814 & $-3.39\pm0.05$ & $0.33\pm0.04$ & $0.61\pm0.07$ & $-0.33\pm0.03$ & $-0.01\pm0.04$ & EMPS \\
     J0908 & $-3.67\pm0.06$ & $0.17\pm0.07$ & $0.35\pm0.04$ & $-0.47\pm0.08$ & $0.35\pm0.08$ & EMPS \\
     J1037 & $-2.50\pm0.05$ & $0.30\pm0.03$ & $0.12\pm0.03$ & $-0.24\pm0.09$ & $0.34\pm0.09$ & CEMPS \\
     J1317 & $-2.37\pm0.05$ & $0.30\pm0.06$ & $0.24\pm0.05$ & $-0.09\pm0.11$ & $0.69\pm0.06$ & VMPS \\
     J1650 & $-2.17\pm0.05$ & $0.13\pm0.07$ & $0.35\pm0.04$ & $-0.22\pm0.10$ & $0.73\pm0.10$ & VMPS \\
     J2242 & $-3.40\pm0.08$ & $0.24\pm0.04$ & $0.71\pm0.14$ & $0.01\pm0.26$ & $0.89\pm0.36$ & EMPS \\
     \hline
    \end{tabular}
    \label{table:jeong_stars}
\end{table*}

\begin{figure}
    \centering
    \includegraphics[width=8.5cm]{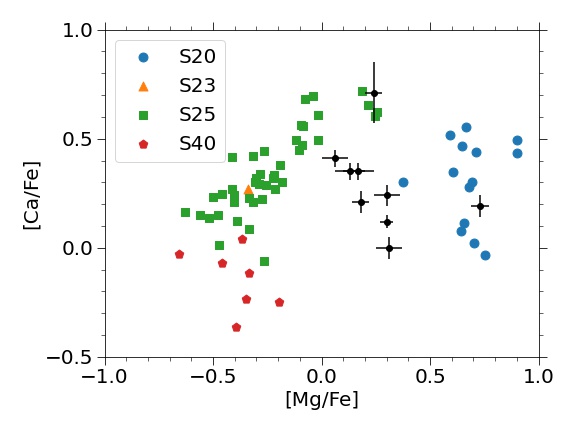}
    \caption{[Ca/Fe] against [Mg/Fe] for the EMPS analyzed in \cite{Jeong2023} (black circle with error bars). Other data points are the theoretical models reported in this work and \cite{Leung2023Jet1}.}
    \label{fig:jeong_CaFe}
\end{figure}

\begin{figure}
    \centering
    \includegraphics[width=8.5cm]{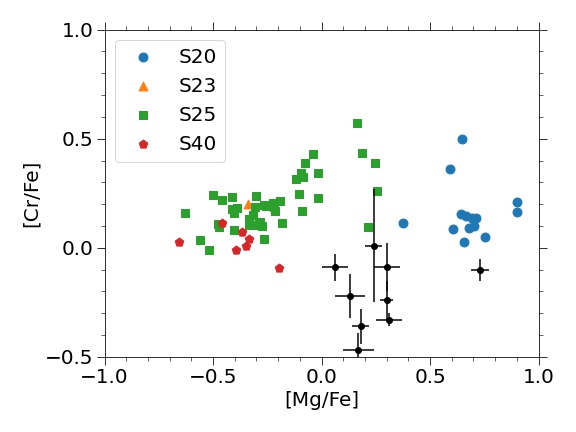}
    \caption{Same as Figure \ref{fig:jeong_CaFe} but for [Cr/Fe] against [Mg/Fe].}
    \label{fig:jeong_CrFe}
\end{figure}

\begin{figure}
    \centering
    \includegraphics[width=8.5cm]{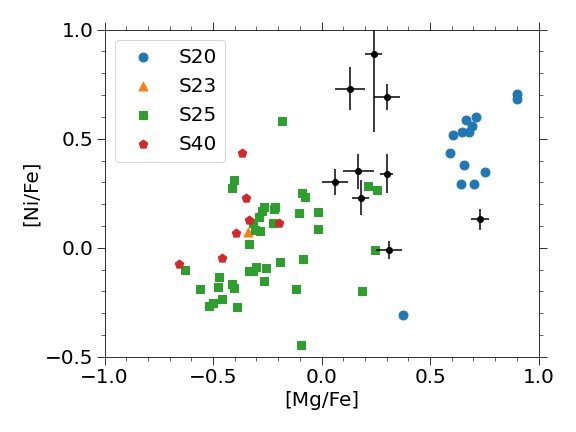}
    \caption{Same as Figure \ref{fig:jeong_CaFe} but for [Ni/Fe] against [Mg/Fe].}
    \label{fig:jeong_NiFe}
\end{figure}

In Figure \ref{fig:jeong_CaFe}, we plot [Ca/Fe] against [Mg/Fe]. Mg is synthesized by C burning and its mass depends on the C/O ratio, which depends on the progenitor mass as is the case of Ne. Thus, this ratio is good for distinguishing the lower mass stars (20 $M_{\odot}$) and the higher mass stars (25 and 40 $M_{\odot}$). Different jet deposition schemes provide such a narrow range of [Ca/Fe] variation as $\sim 0 - 0.5$. The observed stellar data are populated between the lower and higher mass groups of stars. This suggests that the abundances in these observed stars were influenced by the produces of stars in the mass range of $20 - 25~M_{\odot}$ (or the mixture of the two classes of explosion). However, one particular EMPS, J2242, has an exceptionally high [Ca/Fe]. None of our models can reproduce such a high value. On the other hand, the EMPS J0010, is likely to be solely influenced by a 20 $M_{\odot}$ star explosion. 

Figure \ref{fig:jeong_CrFe} is similar to Figure \ref{fig:jeong_CaFe} but for [Cr/Fe] against [Mg/Fe]. Our models show little variation of [Cr/Fe] across different progenitor and explosion models. Most metal-poor stars selected have [Cr/Fe] similar to the theoretical models, except some has lower value by 0.25 dex. 

Figure \ref{fig:jeong_NiFe} is similar to Figure \ref{fig:jeong_CaFe} but for [Ni/Fe] against [Mg/Fe]. There is a clear distinction between the two classes of models. The lower mass models produce [Ni/Fe] $\gtrsim$ 0.3, while the higher mass models produce [Ni/Fe] $\lesssim$ 0.3. This is related to the initial compactness of the pre-explosion progenitor. A lower mass star has a higher central density, thus undergoing more electron capture. Therefore, the explosion produces more neutron-rich spicies, i.e., $^{58}$Ni. Some EMPSs again locate between the two groups, further suggesting the mixed background of massive stars. Some EMPSs exhibit very high [Ni/Fe], e.g., J1317, J1650 and J2242. Since Ni and Fe are synthesized similarly in the inner part of the ejecta, while Cr is produced at further interior part. The high [Ni/Fe] could imply different explosion mechanisms.

\subsection{Metal Poor Star J1010+2358}

The Galactic halo star J1010+2358 was discovered by the LAMOST survey \citep[Large Sky Area Multi-Object Fiber Spectroscopic Telescope;][]{Zhao2006, Zhao2012}, and confirmed to be a very metal poor star (VMPS) J1010+2358 with [Fe/H] $= -2.42$ \citep{Xing2023}. It has an unusually low [Mg/Fe] $\sim -0.66$, which makes it distinctive from typical Galactic halo stars. In the discovery paper, element ratios against Fe, including Mg, Si, Ca, Ti, Cr, Mn, Fe, Co and Ni are reported. The low [Mg/Fe] and [Mn/Mg] suggest less contributions from Type Ia supernovae and canonical core-collapse supernovae. A high mass CCSN produces too high [Mn/Fe] and [Co/Fe] which are less compatible with the observed abundances. They further show that the yield patterns are compatible with the 260 $M_{\odot}$ pair-instability supernovae in \cite{Heger2002}. A recent analysis from \cite{Thibodeaux2024LamostJ1010} favours the origin from low-mass CCSNe. The late-time high-resolution VLT/UVES spectrua where 3D effects and non-LTE effects are accounted also suggest a high C and Al abundances that rule out the pair-instability origin \citep{Skuladottir2024LAMOST}.

In \cite{Xing2023}, the spherical explosion models are compared with this VMP star. As our models suggest that the jet-driven aspherical can lead to distinctive chemical abundance patterns, we extend the search to our jet-driven supernova models. We use the measured elements as the constraints, and look for the best-fit models by using the $\chi$-squared fitting.

\begin{table}[]
    \caption{The $\chi$-squared fitting results and the best-fit models for LAMOST J1010+2358 from various dervied abundances in the literature. "Source" corresponds to the literature work we used for comparing with this VMPS, with "a", "b", "c" standing for \cite{Xing2023}, \cite{Thibodeaux2024LamostJ1010} and 
    \cite{Skuladottir2024LAMOST} respectively.}
    \centering
    \begin{tabular}{c c c c c}
    \hline
        Series & Source & Best-model & $\chi^2$ & best \\ \hline
        20 & a & S20-0250-1000-15 & 64.38 & \\
        25 & a & S25-2000-1000-30 & 25.15 & \checkmark \\
        40 & a & S40-1000-2000-15 & 27.18 & \\
        Spherical & a & 25 ($E_{\rm exp,51} = 5$) & 32.39 & \\ \hline
        20 & b & S20-0250-1000-15 & 130.20 & \\
        25 & b & S25-2000-0500-30 & 57.64 & \\
        40 & b & S40-1000-0500-15 & 33.12 & \checkmark \\
        Spherical & b & 25 ($E_{\rm exp,51} = 10$) & 58.40 & \\ \hline
        20 & c & S20-0250-1000-15 & 118.84 & \\
        25 & c & S25b-1000-1000-30 & 69.58 & \\ 
        40 & c & S40-1000-1000-15 & 46.76 & \checkmark \\ 
        Spherical & c & 30 ($E_{\rm exp,51} = 20$) & 90.78 & \\
        \hline
    \end{tabular}
    \label{tab:compare_LAMOST}
\end{table}

In Table \ref{tab:compare_LAMOST} we show our $\chi$-square fitting results for the LAMOST J1010+2358 using the abundances derived in the three works. The data from \cite{Xing2023} has the closest abundance pattern to our model with the lowest $\chi$-square score. The Model S25-2000-1000-30 is the best fit models. The new data derived in \cite{Thibodeaux2024LamostJ1010} and \cite{Skuladottir2024LAMOST} are showing slightly larger discrepency for the best-fit models. Both set of data suggest 40 $M_{\odot}$ models. The unique features in J1010+2358 strongly exclude the interpretation with the lower mass CCSN (20 $M_{\odot}$) or direct 1D spherical explosions. We also show in Table \ref{table:LAMOST} our comparison results. It shows that the parameters close to the characteristic model can already provide a close fit to this object. The best-fit model is found to be $M_{\rm ZAMS} = 25~M_{\odot}$. 

\begin{figure}
    \centering
    \includegraphics[width=8.5cm]{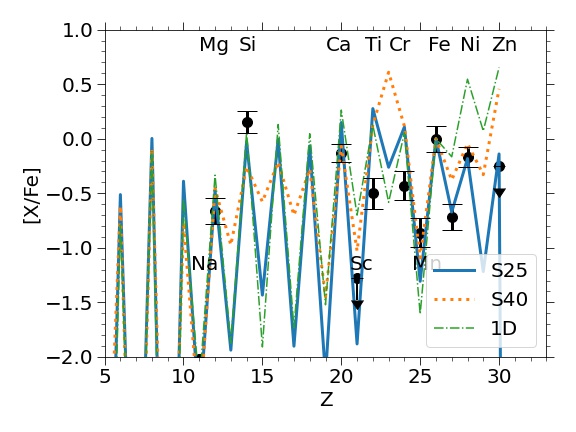}
    \includegraphics[width=8.5cm]{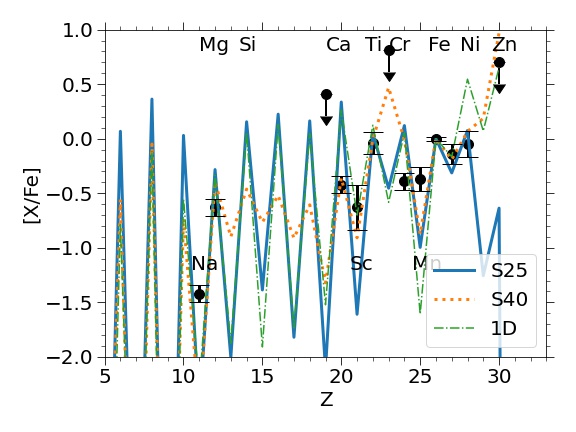}
    \includegraphics[width=8.5cm]{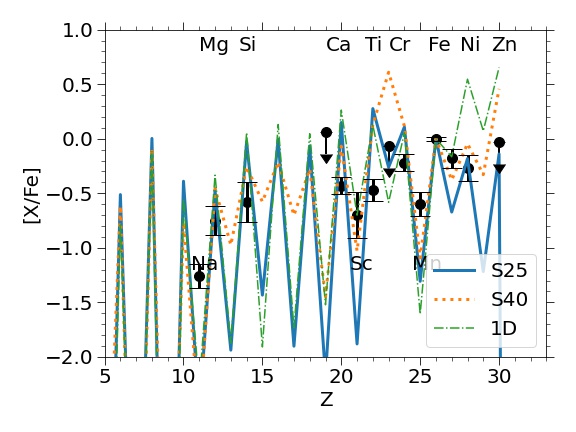}

    \caption{(top panel) The chemical abundance of J1010$+$2358 \citep[circles with error bars, ][]{Xing2023} and the best-fit models searched from the jet-driven supernova models with $M_{\rm prog} = 25, 40~M_{\odot}$ and spherical explosion models. (middle panel) Same as the top panel, but with the data from \citep{Thibodeaux2024LamostJ1010}. (bottom panel) Same as the top panel, but with the data from \citep{Skuladottir2024LAMOST}. 
    }
    \label{fig:LAMOST_yield}
\end{figure}

In Figure \ref{fig:LAMOST_yield} we show our comparisons. In the top panel, we plot the data points from the LAMOST J1010+2358 interpreted in \cite{Xing2023}, together with the best models from our 25-, 40-$M_{\odot}$ and spherical model series. The models represent those of the minimum $\chi$-square. For the 1D model, it can fit the Mg and Ca. But its prediction of Ti, Cr, Ni and Zn are too high to be compatible. The 25 $M_{\odot}$ is the best models among all interpretations and mass sequences. The pattern essentially mimics all elements, with an exception for Ti and Cr. The 40 $M_{\odot}$ has a similar pattern but has too high [Ti/Fe], [Cr/Fe] and [Zn/Fe] to match the VMPS. We remarked that Zn/Fe in the model is sensitive to the mass cut, so that the comparison of Zn/Fe cannot rule out the  model of Zn/Fe.

In the middle panel, we plot the data points similar to the top panel but for the dataset derived in \cite{Thibodeaux2024LamostJ1010}. Their data set does not contain Si, a lower Ca and higher Ti, Mn and Co. The dataset also contains values for Sc. The 25 $M_{\odot}$ model cannot match the high [Sc/Fe] and [Mn/Fe] observed in that star. The 40 $M_{\odot}$ model, being the best-fit models among all, can track all the elements except that the [Mn/Fe] in our model is too low to match the close-to-solar value in the VMPS. In general, our model predicts a sub-solar value of [Mn/Fe] because Mn is in general a decay product of $^{55}$Co, which is one of the end-product of $^{22}$Ne. For supernovae exploding from zero-metallicity star, the star does not begin with any observable amount of $^{14}$N and $^{22}$Ne, hence its production of $^{55}$Mn is naturally suppressed. The 1D model has a similar abundance pattern as the 25 $M_{\odot}$, where the [Ca/Fe] and [Cr/Fe] are too high, and [Mn/Fe] is too low which plaques the spherical model to explain the VMPS.

In the bottom panel, we plot the data point similar to the top panel but for the dataset derived in \cite{Skuladottir2024LAMOST}. The dataset features a more conserved [Si/Fe], [Ti/Fe] and [Mn/Fe] compared to \cite{Thibodeaux2024LamostJ1010}, but some, such as [Sc/Fe], [Co/Fe], are similar to that work. The 1D model predicts too high [Mg/Fe], [Si/Fe], [Ca/Fe], [Ti/Fe] and [Ni/Fe] as in previous comparisons. The 25 $M_{\odot}$ model also suffers from similar problems and fails to be the best-fit model, except that it has a more comparable [Ni/Fe] and [Zn/Fe] in that object. The 40 $M_{\odot}$ model is more compatible with LAMOST J1010+2358 due to its lower [Mg/Fe], [Si/Fe] and the Fe-group elements. Still, it predicts the [Ti/Fe] and [Zn/Fe] too high while [Mn/Fe] being too low to be a satisfactory interpretation of this VMPS.

The object is extensively compared in \cite{Xing2023}. The comparison with the Chandrasekhar mass white dwarf yields are used to compare with this VMP star. However, in such an early system, we expect that the contribution from the Chandrasekhar mass white dwarf explosion is low because of the time delay from the formation of the white dwarf until it reaches the ignition mass \citep{Kobayashi2020}. In fact, the sub-Chandrasekhar mass Type Ia supernova yields in \cite{Leung2020subChand} show low or sub-solar [Ti/Fe], [Cr/Fe], [Mn/Fe] and [Co/Fe] in contrast to the Chandrasekhar mass counterpart \citep[see e.g.,][]{Nomoto2017SNIa, Nomoto2018, Leung2018Chand}. The mixing of yields of jet-driven supernovae with the low metallicity sub-Chandrasekhar models can lead to lower mass fraction ratios in Ti and Cr.

Given the strong mass dependence of [Ne/O] as demonstrated in the previous section, the knowledge of spectral data from lines of these two elements will provide important insights for determining the origin of the supernova (1D vs. 2D CCSN, PISN). In particular, a major disparity comes from the stringent constraints of [Ti/Fe] and [Cr/Fe] where our models tend to predict a high value while LAMOST J1010+2358 has a close-to-solar values. The production of Ti and Cr centered around matter with a temperature $\sim 4-5 \times 10^9$ K with a density about $10^6 -10^7$ g cm$^{-3}$. As shown in Figure \ref{fig:Ti_V_plot}, most jet-driven supernova models tend to produce solar to even super-solar [Ti/Fe] and [V/Fe]. In Table \ref{tab:compare_LAMOST} it becomes clear that different spectral modeling and input physics (e.g., NLTE or not) could lead to distinctive interpretations of the chemical abundances. Future accurate and model-independent measurement of these objects will provide further hints on the shock-heating history in the ejecta, and the general progenitor mass of the VMPS.

\begin{table}[]
    \caption{The best-fit models from our numerical models based on each massive star progenitors based on $\chi$-square fitting using the 11 element ratios as constraints. The column ``Explosion'' stands for whether the 2D-jet or the 1D spherical explosion is used. The Model is the best-fit model for that progenitor.}
    \centering
    \begin{tabular}{c c c c}
    \hline
         Explosion & $M_{\rm ZAMS}$ & Model & $\chi^2$ \\ \hline
         Jet & 20 & S20-0250-1000-15 & 64.38 \\ 
         Jet & 25 & S25-2000-1000-30 & 25.15 \\
         Jet & 40 & S40-1000-2000-15 & 27.18 \\ 
         Spherical & 25 & $E_{\rm exp} = 5\times10^{51}$ erg & 32.38 \\ \hline 
    \end{tabular}
    \label{table:LAMOST}
\end{table}

\section{Conclusion}

In this article, we compute nucleosynthesis yields of our jet-driven supernova models using multi-dimensional simulations. We calculate explosions of massive stars from 20 -- 40 $M_{\odot}$ and associated explosive nucleosynthesis. We study how the explosion hydrodynamics and nucleosynthesis depend on the progenitor properties (mass, mass cut, and dimensionality). We compare the ejecta observable, including chemical abundances with available metal-poor stars, supernova, and metal-poor galaxies. Some of our models are consistent with the recently observed extremely metal-poor galaxies (EMPGs). Below we list our findings of this work:

\begin{enumerate}
    \item The inner mass cut and the initial black hole mass strongly change the [Fe/O] ratio of the ejecta and the ejecta mass.
    \item Asphericity of the explosion provides a reasonable energy budget to match the observed high velocity, and the thermodynamics history is distinct from the spherical counterpart.
    \item The spread of the explosion energetics can explain the diversity of the observed supernova ejecta mass and the $^{56}$Ni distribution. Some parameter space featuring the extremely high $^{56}$Ni mass ($>1~M_{\odot}$) and high ejecta mass mass ($\sim 10~M_{\odot}$) remains to be solved.
    \item The lower mass $(\sim 25~M_{\odot})$ models can produce the Ti-V relation being consistent with that from the metal-poor stars in the stellar survey in the literature. 
    \item The extremely metal-poor galaxies (EMPGs) can be characterized by their [Ne/O] which is sensitive to the progenitor mass. A high [Ne/O] corresponds to a lower progenitor mass and vice versa.
    
    \item Despite the extremely metal-poor nature, i.e., even without contribution of Fe from Type Ia supernovae, the abundance ratios [(Si, S, Ar, Ca, Fe)/O] of these EMPGs are as high as the solar values, which can be matched with single or a few jet-driven supernova explosions.
    \item In the energetic jet-driven supernovae, despite explosive O-burning converts O to Si, S, A, and Ca, [(Si, S, Ar, Ca)/O] do not exceed the solar ratios. This is consistent with the observed ratios in EMPGs.
    \item The Very Metal Poor star J1010+2358 in the Galactic Halo, poor in [Mg/Fe] and [Mn/Fe] can be well matched with our jet-driven supernova of the 25 $M_{\odot}$ progenitor.
    \item Metal-poor stars tend to exhibit the abundance pattern of a mixture of multiple supernova explosions instead of a single one. 
\end{enumerate}

\section*{Acknowledgement}

We thank Frank Timmes for the open-source subroutines of the Helmholtz equation of state and the torch nuclear reaction network.
This material is based upon work supported by the National Science Foundation under Grant AST-2316807.
K.N. acknowledges support by World Premier International Research Center Initiative (WPI), and JSPS KAKENHI Grant Numbers JP20K04024, JP21H04499, and JP23K03452. 

\software{  Numpy \citep{Numpy},
            Matplotlib \citep{Matplotlib},
            Pandas \citep{Pandas}
          }

\newpage

\appendix

\begin{table}[h]
    \centering
    \caption{The masses of Fe and $^{56}$Ni of the models in S20 series and the elemental ratios [Z/O] of other major elements, defined in Equation (1). Masses are in units of the solar mass. See also Table \ref{table:S20models} -- \ref{table:S25cmodels} for M(O) and other elements of this model series.}
    \begin{tabular}{c c c c c c c c c c c c}
        \hline
        Model & $M$(Fe) & $M$($^{56}$Ni) & [C/O] & [Mg/O] & [Ca/O] & [Ti/O] & [Cr/O] & [Mn/O] & [Co/O] & [Ni/O] & [Zn/O] \\ \hline
S20-0250-1000-15  &  0.038 & 0.035 & -0.56 & -0.15 & -1.00 & -0.48 & -0.84 & -1.80 & -0.98 & -0.71 & -0.93 \\
S20-0500- 500-15  &  0.037 & 0.035 & -0.61 & -0.15 & -0.82 & -0.52 & -0.75 & -1.51 & -0.89 & -0.56 & -1.16 \\
S20-0500-1000-15  &  0.045 & 0.042 & -0.62 & -0.15 & -0.71 & -0.62 & -0.64 & -1.00 & -0.80 & -0.50 & -1.15 \\
S20-0500-2000-15  &  0.046 & 0.043 & -0.63 & -0.15 & -0.69 & -0.71 & -0.78 & -1.20 & -0.78 & -0.43 & -1.12 \\
S20-1000-0250-15  &  0.037 & 0.034 & -0.63 & -0.14 & -0.93 & -0.66 & -0.85 & -1.60 & -1.01 & -0.55 & -1.22 \\
S20-1000-0500-15  &  0.047 & 0.044 & -0.64 & -0.15 & -0.55 & -0.76 & -0.74 & -0.83 & -0.80 & -0.30 & -1.21 \\
S20-1000-1000-15  &  0.029 & 0.027 & -0.64 & -0.15 & -0.61 & -1.08 & -0.88 & -0.90 & -0.87 & -0.35 & -1.88 \\
S20-1000-2000-15  &  0.030 & 0.027 & -0.65 & -0.15 & -0.56 & -0.97 & -0.84 & -0.88 & -0.87 & -0.37 & -2.00 \\
S20-1000-4000-15  &  0.047 & 0.043 & -0.65 & -0.15 & -0.54 & -0.69 & -0.70 & -0.69 & -0.73 & -0.28 & -1.67 \\
S20-2000-0500-15  &  0.058 & 0.053 & -0.65 & -0.15 & -0.41 & -0.74 & -0.67 & -0.69 & -0.57 & -0.24 & -1.12 \\
S20-2000-1000-15  &  0.047 & 0.044 & -0.66 & -0.16 & -0.43 & -0.87 & -0.73 & -0.78 & -0.64 & -0.27 & -1.54 \\
S20-2000-2000-15  &  0.052 & 0.047 & -0.67 & -0.16 & -0.27 & -0.80 & -0.68 & -0.55 & -0.42 & -0.23 & -1.46 \\
S20-4000-1000-15  &  0.064 & 0.053 & -0.66 & -0.15 & -0.23 & -0.63 & -0.38 & -0.22 & -0.48 & -0.31 & -1.65 \\
S20-4000-2000-15  &  0.057 & 0.039 & -0.66 & -0.15 & -0.33 & -0.71 & -0.30 & -0.07 & -0.35 & -0.27 & -1.69 \\ \hline
    \end{tabular}

    \label{table:yields2_S20}
\end{table}

\begin{table}[h]
    \centering
    \caption{Same as Table \ref{table:yields2_S20} but for the models in Series S25a. See also Table \ref{table:S20models} -- \ref{table:S25cmodels} for M(O) and other elements of this model series.}
    \begin{tabular}{c c c c c c c c c c c c}
        \hline
        Model & $M$(Fe) & $M$($^{56}$Ni) & [C/O] & [Mg/O] & [Ca/O] & [Ti/O] & [Cr/O] & [Mn/O] & [Co/O] & [Ni/O] & [Zn/O] \\ \hline
S25a-0250-1000-15  &  0.110 & 0.105 & -0.25 & -0.65 &  0.14 &  0.30 &  0.13 & -1.17 & -0.93 & -0.13 & -0.05 \\
S25a-0250-2000-15  &  0.014 & 0.014 & -0.22 & -0.72 &  0.14 & -0.54 & -0.32 & -1.63 & -2.24 & -1.84 & -2.46 \\
S25a-0500-0500-15  &  0.100 & 0.096 & -0.26 & -0.68 & -0.17 &  0.02 & -0.11 & -1.53 & -1.16 & -0.56 & -0.41 \\
S25a-0500-1000-15  &  0.151 & 0.144 & -0.34 & -0.64 &  0.06 &  0.11 &  0.04 & -1.36 & -0.81 & -0.42 & -0.52 \\
S25a-0500-2000-15  &  0.060 & 0.058 & -0.33 & -0.67 & -0.01 & -0.44 & -0.23 & -1.63 & -1.40 & -1.02 & -0.51 \\
S25a-0500-4000-15  &  0.167 & 0.160 & -0.39 & -0.60 &  0.04 & -0.03 & -0.04 & -1.35 & -0.68 & -0.38 & -0.61 \\
S25a-1000-0250-15  &  0.042 & 0.040 & -0.22 & -0.68 & -0.47 & -0.01 & -0.37 & -1.93 & -1.20 & -0.56 & -0.40 \\
S25a-1000-0500-15  &  0.149 & 0.142 & -0.43 & -0.57 & -0.02 & -0.10 & -0.12 & -1.46 & -0.75 & -0.40 & -0.85 \\
S25a-1000-1000-15  &  0.206 & 0.197 & -0.45 & -0.54 &  0.07 & -0.07 & -0.01 & -1.25 & -0.53 & -0.13 &  0.70 \\
S25a-1000-1000-30  &  0.262 & 0.253 & -0.47 & -0.51 &  0.10 &  0.08 & -0.03 & -1.45 & -0.25 &  0.20 &  1.28 \\
S25a-1000-2000-15  &  0.213 & 0.204 & -0.46 & -0.52 &  0.04 & -0.12 & -0.06 & -1.30 & -0.55 & -0.17 & -0.62 \\
S25a-1000-2000-30  &  0.353 & 0.341 & -0.48 & -0.51 &  0.15 &  0.11 &  0.00 & -1.47 & -1.05 & -0.25 & -0.10 \\
S25a-1000-4000-15  &  0.056 & 0.054 & -0.45 & -0.50 & -0.15 & -0.63 & -0.37 & -0.92 & -0.90 & -0.76 & -1.81 \\
S25a-2000-0250-15  &  0.216 & 0.207 & -0.47 & -0.53 & -0.11 & -0.09 & -0.09 & -1.40 & -0.68 & -0.30 & -0.54 \\
S25a-2000-0500-15  &  0.247 & 0.236 & -0.50 & -0.49 &  0.04 & -0.14 & -0.06 & -0.90 & -0.41 & -0.06 & -0.64 \\
S25a-2000-1000-15  &  0.235 & 0.225 & -0.51 & -0.48 &  0.02 & -0.16 & -0.10 & -1.32 & -0.50 & -0.03 & -0.63 \\
S25a-2000-1000-30  &  0.367 & 0.352 & -0.51 & -0.48 &  0.15 &  0.27 &  0.10 & -1.31 & -0.67 & -0.18 & -0.14 \\
S25a-2000-2000-15  &  0.198 & 0.184 & -0.52 & -0.47 &  0.02 & -0.28 & -0.17 & -0.68 & -0.09 &  0.30 &  0.94 \\
S25a-2000-2000-30  &  0.427 & 0.410 & -0.52 & -0.50 &  0.21 &  0.17 &  0.10 & -1.34 & -0.55 & -0.12 & -0.24 \\
S25a-4000-0500-15  &  0.280 & 0.269 & -0.54 & -0.44 &  0.14 & -0.06 & -0.03 & -1.27 & -0.52 & -0.01 & -0.57 \\
S25a-4000-1000-15  &  0.252 & 0.242 & -0.55 & -0.46 &  0.25 & -0.16 &  0.00 & -0.91 & -0.35 & -0.00 & -0.79 \\
S25a-4000-2000-15  &  0.332 & 0.311 & -0.54 & -0.47 &  0.36 & -0.11 &  0.12 & -0.08 & -0.09 &  0.21 & -0.60 \\ \hline
    \end{tabular}

    \label{table:yields2_S25a}
\end{table}

\begin{table}[h]
    \centering
    \caption{Same as Table \ref{table:yields2_S20} but for the models in Series S25b. See also Table \ref{table:S20models} -- \ref{table:S25cmodels} for M(O) and other elements of this model series.}
    \begin{tabular}{c c c c c c c c c c c c}
        \hline
        Model & $M$(Fe) & $M$($^{56}$Ni) & [C/O] & [Mg/O] & [Ca/O] & [Ti/O] & [Cr/O] & [Mn/O] & [Co/O] & [Ni/O] & [Zn/O] \\ \hline
S25b-0500-1000-15  &  0.080 & 0.077 & -0.28 & -0.65 &  0.09 &  0.18 &  0.10 & -1.26 & -0.72 & -0.40 & -0.54 \\
S25b-0500-2000-15  &  0.024 & 0.023 & -0.24 & -0.71 &  0.03 & -0.50 & -0.24 & -1.49 & -1.67 & -1.39 & -1.83 \\
S25b-1000-0500-15  &  0.058 & 0.056 & -0.34 & -0.62 & -0.01 & -0.34 & -0.19 & -1.56 & -0.85 & -0.69 & -1.46 \\
S25b-1000-1000-15  &  0.108 & 0.103 & -0.39 & -0.59 &  0.03 & -0.11 & -0.05 & -1.53 & -0.63 & -0.38 & -0.85 \\
S25b-1000-2000-15  &  0.121 & 0.116 & -0.40 & -0.56 &  0.04 & -0.16 & -0.07 & -1.29 & -0.64 & -0.18 & -0.74 \\
S25b-1000-4000-15  &  0.043 & 0.041 & -0.44 & -0.50 & -0.14 & -0.71 & -0.50 & -1.25 & -0.90 & -0.50 & -1.96 \\
S25b-2000-0500-15  &  0.132 & 0.127 & -0.46 & -0.52 &  0.02 & -0.18 & -0.09 & -1.40 & -0.57 & -0.18 & -0.95 \\
S25b-2000-1000-15  &  0.139 & 0.134 & -0.48 & -0.51 &  0.05 & -0.23 & -0.10 & -1.22 & -0.52 & -0.11 & -0.96 \\
S25b-2000-2000-15  &  0.143 & 0.137 & -0.48 & -0.49 & -0.01 & -0.30 & -0.11 & -0.85 & -0.47 & -0.09 & -0.91 \\
S25b-4000-1000-15  &  0.119 & 0.115 & -0.51 & -0.48 &  0.17 & -0.34 & -0.07 & -1.42 & -0.73 & -0.44 & -1.59 \\
S25b-4000-2000-15  &  0.110 & 0.102 & -0.52 & -0.50 &  0.26 & -0.34 & -0.03 & -0.29 & -0.51 & -0.19 & -2.12 \\ \hline
    \end{tabular}

    \label{table:yields2_S25b}
\end{table}

\begin{table}[h]
    \centering
    \caption{Same as Table \ref{table:yields2_S20} but for the models in Series S25c. See also Table \ref{table:S20models} -- \ref{table:S25cmodels} for M(O) and other elements of this model series.}
    \begin{tabular}{c c c c c c c c c c c c}
        \hline
        Model & $M$(Fe) & $M$($^{56}$Ni) & [C/O] & [Mg/O] & [Ca/O] & [Ti/O] & [Cr/O] & [Mn/O] & [Co/O] & [Ni/O] & [Zn/O] \\ \hline
S25c-0500-1000-15  &  0.070 & 0.067 & -0.30 & -0.61 & -0.12 &  0.04 & -0.04 & -1.29 & -0.62 & -0.26 & -0.59 \\
S25c-0500-2000-15  &  0.062 & 0.060 & -0.27 & -0.67 &  0.07 & -0.33 & -0.20 & -1.60 & -0.84 & -0.46 & -1.07 \\
S25c-1000-0500-15  &  0.083 & 0.080 & -0.27 & -0.65 &  0.03 & -0.04 & -0.00 & -1.42 & -0.62 & -0.40 & -0.84 \\
S25c-1000-1000-15  &  0.078 & 0.074 & -0.30 & -0.65 & -0.03 & -0.31 & -0.24 & -1.36 & -0.68 & -0.29 & -1.00 \\
S25c-1000-2000-15  &  0.066 & 0.064 & -0.36 & -0.61 & -0.10 & -0.51 & -0.36 & -1.45 & -0.92 & -0.50 & -1.16 \\
S25c-1000-4000-15  &  0.059 & 0.057 & -0.43 & -0.53 & -0.08 & -0.67 & -0.64 & -1.44 & -0.82 & -0.45 & -1.64 \\
S25c-2000-0500-15  &  0.111 & 0.107 & -0.38 & -0.59 & -0.02 & -0.31 & -0.19 & -1.52 & -0.83 & -0.46 & -1.04 \\
S25c-2000-1000-15  &  0.120 & 0.114 & -0.44 & -0.55 &  0.00 & -0.35 & -0.20 & -1.10 & -0.57 & -0.29 & -1.08 \\
S25c-2000-2000-15  &  0.143 & 0.138 & -0.49 & -0.51 &  0.05 & -0.43 & -0.25 & -1.24 & -0.58 & -0.17 & -1.20 \\
S25c-4000-1000-15  &  0.076 & 0.074 & -0.49 & -0.51 &  0.02 & -0.55 & -0.26 & -1.40 & -1.03 & -0.89 & -2.29 \\
S25c-4000-2000-15  &  0.127 & 0.122 & -0.50 & -0.50 &  0.12 & -0.46 & -0.14 & -0.76 & -0.72 & -0.32 & -1.48 \\
\hline
    \end{tabular}

    \label{table:yields2_S25c}
\end{table}

\section{Nucleosynthetic Yields of Jet-Driven Supernova Models}

In Section \ref{sec:models} we present in details how the dynamics and nucleosythesis yields depend on the dimension, jet energetics and mass cut, and focus on some key elements including [Ne/O], [Si/O], [S/O], [Fe/O] which are commonly observed in EMPGs. Here, we present yields of other important elemental pairs for further comparison with available observation data. In Table \ref{table:yields2_S20} -- \ref{table:yields2_S25c}, we list the results for the for the model series of the S20, S25a, S25b and S25c, respectively.

\section{Nucleosynthetic Yields of Spherical Models}

In Section \ref{sec:models} we described how the mixing-fallback mechanism of the spherical explosion model can mimick the jet-driven supernovae. Here we present the total nucleosynthesic yields of the massive star explosions assuming spherical explosion of various masses (20, 25, and 40 $M_{\odot}$) and explosion energies ($10^{51}$ and $10^{52}$ erg). The entire envelope is assumed to be ejected without mixing-and-fallback. In Tables \ref{table:yield_isotopes} -- \ref{table:yield_elements} we list the nucleosynthetic yields of these explosion models. 

\begin{longtable}[h]{c c c c c c c}
 \caption{The isotope yields table for spherical explosion models. Masses are in units of $M_{\odot}$ and energies are in unit of $10^{51}$ erg.\label{table:yield_isotopes}} \\
 
 Isotope & $M=20, E=1$ & $M=25, E=1$ & $M=40, E=1$ & $M=20, E=10$ & $M=25, E=10$ & $M=40, E=10$ \\
 \hline
 \endfirsthead

 \multicolumn{7}{c}{\textit{Continuation of Table \ref{table:yield_isotopes}.}} \\
 Isotope & $M=20, E=1$ & $M=25, E=1$ & $M=40, E=1$ & $M=20, E=10$ & $M=25, E=10$ & $M=40, E=10$ \\
 \hline
 \endhead

 \hline
 \endfoot

 \hline
 \endlastfoot

 $^{12}$C & $2.12 \times 10^{-1}$ & $3.6 \times 10^{-1}$ & $4.33 \times 10^{-1}$ & $1.89 \times 10^{-1}$ & $2.73 \times 10^{-1}$ & $4.32 \times 10^{-1}$ \\
 $^{13}$C & $2.34 \times 10^{-9}$ & $1.25 \times 10^{-9}$ & $1.32 \times 10^{-9}$ & $1.39 \times 10^{-9}$ & $6.99 \times 10^{-8}$ & $5.57 \times 10^{-9}$ \\
 $^{14}$N & $3.10 \times 10^{-7}$ & $5.47 \times 10^{-7}$ & $9.45 \times 10^{-8}$ & $2.56 \times 10^{-7}$ & $1.61 \times 10^{-6}$ & $3.63 \times 10^{-7}$ \\
 $^{15}$N & $7.5 \times 10^{-7}$ & $3.74 \times 10^{-6}$ & $9.48 \times 10^{-6}$ & $1.40 \times 10^{-6}$ & $4.62 \times 10^{-6}$ & $1.17 \times 10^{-5}$ \\
 $^{16}$O & 2.15 & 2.84 & 8.24 & 2.02 & 2.26 & 7.50 \\
 $^{17}$O & $5.19 \times 10^{-8}$ & $3.73 \times 10^{-9}$ & $1.54 \times 10^{-9}$ & $2.85 \times 10^{-9}$ & $1.8 \times 10^{-8}$ & $3.27 \times 10^{-9}$ \\
 $^{18}$O & $5.77 \times 10^{-11}$ & $4.32 \times 10^{-9}$ & $5.31 \times 10^{-12}$ & $1.93 \times 10^{-7}$ & $2.96 \times 10^{-8}$ & $3.27 \times 10^{-11}$ \\
 $^{19}$F & $1.89 \times 10^{-10}$ & $1.93 \times 10^{-10}$ & $8.97 \times 10^{-11}$ & $1.61 \times 10^{-10}$ & $6.15 \times 10^{-10}$ & $2.66 \times 10^{-10}$ \\
 $^{20}$Ne & $9.19 \times 10^{-1}$ & $5.18 \times 10^{-1}$ & $3.11 \times 10^{-1}$ & $7.29 \times 10^{-1}$ & $2.53 \times 10^{-1}$ & $3.25 \times 10^{-1}$ \\
 $^{21}$Ne & $3.35 \times 10^{-7}$ & $6.66 \times 10^{-7}$ & $3.19 \times 10^{-7}$ & $4.70 \times 10^{-7}$ & $1.32 \times 10^{-5}$ & $8.16 \times 10^{-7}$ \\
 $^{22}$Ne & $4.38 \times 10^{-7}$ & $4.00 \times 10^{-6}$ & $7.10 \times 10^{-8}$ & $6.25 \times 10^{-6}$ & $5.71 \times 10^{-6}$ & $6.41 \times 10^{-6}$ \\
 $^{23}$Na & $2.22 \times 10^{-5}$ & $2.90 \times 10^{-5}$ & $1.40 \times 10^{-6}$ & $3.56 \times 10^{-5}$ & $5.64 \times 10^{-5}$ & $5.65 \times 10^{-6}$ \\
 $^{24}$Mg & $1.45 \times 10^{-1}$ & $1.10 \times 10^{-1}$ & $4.63 \times 10^{-1}$ & $1.39 \times 10^{-1}$ & $9.11 \times 10^{-2}$ & $3.89 \times 10^{-1}$ \\
 $^{25}$Mg & $1.57 \times 10^{-6}$ & $2.26 \times 10^{-6}$ & $7.14 \times 10^{-7}$ & $5.14 \times 10^{-6}$ & $1.38 \times 10^{-5}$ & $2.91 \times 10^{-6}$ \\
 $^{26}$Mg & $3.10 \times 10^{-6}$ & $1.20 \times 10^{-5}$ & $8.24 \times 10^{-7}$ & $2.30 \times 10^{-5}$ & $2.48 \times 10^{-5}$ & $1.53 \times 10^{-5}$ \\
 $^{26}$Al & $4.66 \times 10^{-28}$ & $1.78 \times 10^{-26}$ & $9.52 \times 10^{-28}$ & $5.68 \times 10^{-27}$ & $1.75 \times 10^{-26}$ & $9.52 \times 10^{-28}$ \\
 $^{27}$Al & $8.95 \times 10^{-5}$ & $1.79 \times 10^{-4}$ & $1.10 \times 10^{-4}$ & $2.20 \times 10^{-4}$ & $3.38 \times 10^{-4}$ & $1.76 \times 10^{-4}$ \\
 $^{28}$Si & $9.20 \times 10^{-2}$ & $3.9 \times 10^{-1}$ & $9.21 \times 10^{-1}$ & $1.45 \times 10^{-1}$ & $4.15 \times 10^{-1}$ & $9.70 \times 10^{-1}$ \\
 $^{29}$Si & $2.60 \times 10^{-5}$ & $6.23 \times 10^{-5}$ & $1.66 \times 10^{-4}$ & $8.17 \times 10^{-5}$ & $1.61 \times 10^{-4}$ & $2.6 \times 10^{-4}$ \\
 $^{30}$Si & $2.69 \times 10^{-5}$ & $4.20 \times 10^{-5}$ & $4.20 \times 10^{-5}$ & $1.50 \times 10^{-4}$ & $2.18 \times 10^{-4}$ & $1.19 \times 10^{-4}$ \\
 $^{31}$P & $3.84 \times 10^{-5}$ & $1.0 \times 10^{-4}$ & $1.27 \times 10^{-4}$ & $1.14 \times 10^{-4}$ & $1.85 \times 10^{-4}$ & $2.64 \times 10^{-4}$ \\
 $^{32}$S & $5.67 \times 10^{-2}$ & $1.76 \times 10^{-1}$ & $4.85 \times 10^{-1}$ & $6.58 \times 10^{-2}$ & $2.32 \times 10^{-1}$ & $5.31 \times 10^{-1}$ \\
 $^{33}$S & $3.57 \times 10^{-5}$ & $4.87 \times 10^{-5}$ & $5.9 \times 10^{-5}$ & $5.95 \times 10^{-5}$ & $9.19 \times 10^{-5}$ & $9.66 \times 10^{-5}$ \\
 $^{34}$S & $2.50 \times 10^{-5}$ & $2.70 \times 10^{-5}$ & $9.90 \times 10^{-6}$ & $1.55 \times 10^{-4}$ & $2.77 \times 10^{-4}$ & $7.27 \times 10^{-5}$ \\
 $^{36}$S & $1.82 \times 10^{-11}$ & $4.90 \times 10^{-12}$ & $4.51 \times 10^{-12}$ & $8.59 \times 10^{-12}$ & $7.50 \times 10^{-12}$ & $5.69 \times 10^{-12}$ \\
 $^{35}$Cl & $1.56 \times 10^{-5}$ & $3.29 \times 10^{-5}$ & $5.52 \times 10^{-5}$ & $1.70 \times 10^{-4}$ & $1.80 \times 10^{-4}$ & $1.37 \times 10^{-4}$ \\
 $^{37}$Cl & $2.20 \times 10^{-5}$ & $1.50 \times 10^{-5}$ & $6.42 \times 10^{-6}$ & $1.10 \times 10^{-5}$ & $1.30 \times 10^{-5}$ & $2.29 \times 10^{-5}$ \\
 $^{36}$Ar & $1.26 \times 10^{-2}$ & $3.56 \times 10^{-2}$ & $9.83 \times 10^{-2}$ & $1.37 \times 10^{-2}$ & $5.15 \times 10^{-2}$ & $1.9 \times 10^{-1}$ \\
 $^{38}$Ar & $1.3 \times 10^{-5}$ & $1.0 \times 10^{-5}$ & $2.66 \times 10^{-6}$ & $3.17 \times 10^{-5}$ & $5.78 \times 10^{-5}$ & $1.95 \times 10^{-5}$ \\
 $^{40}$Ar & $2.36 \times 10^{-14}$ & $5.63 \times 10^{-15}$ & $8.58 \times 10^{-15}$ & $7.44 \times 10^{-15}$ & $4.66 \times 10^{-15}$ & $2.80 \times 10^{-15}$ \\
 $^{39}$K & $1.53 \times 10^{-5}$ & $3.18 \times 10^{-5}$ & $2.86 \times 10^{-5}$ & $7.20 \times 10^{-5}$ & $1.30 \times 10^{-4}$ & $7.4 \times 10^{-5}$ \\
 $^{40}$K & $3.60 \times 10^{-11}$ & $8.51 \times 10^{-12}$ & $9.61 \times 10^{-12}$ & $8.56 \times 10^{-12}$ & $4.48 \times 10^{-12}$ & $4.24 \times 10^{-12}$ \\
 $^{41}$K & $9.89 \times 10^{-6}$ & $3.53 \times 10^{-6}$ & $1.68 \times 10^{-6}$ & $3.85 \times 10^{-6}$ & $1.55 \times 10^{-6}$ & $7.45 \times 10^{-6}$ \\
 $^{40}$Ca & $1.12 \times 10^{-2}$ & $3.33 \times 10^{-2}$ & $1.00 \times 10^{-1}$ & $1.45 \times 10^{-2}$ & $5.55 \times 10^{-2}$ & $1.10 \times 10^{-1}$ \\
 $^{42}$Ca & $1.46 \times 10^{-6}$ & $5.54 \times 10^{-6}$ & $2.46 \times 10^{-7}$ & $8.68 \times 10^{-6}$ & $1.61 \times 10^{-5}$ & $1.11 \times 10^{-6}$ \\
 $^{43}$Ca & $2.45 \times 10^{-6}$ & $1.49 \times 10^{-6}$ & $7.60 \times 10^{-7}$ & $4.55 \times 10^{-6}$ & $5.15 \times 10^{-6}$ & $2.62 \times 10^{-6}$ \\
 $^{44}$Ca & $3.74 \times 10^{-5}$ & $9.50 \times 10^{-5}$ & $3.12 \times 10^{-4}$ & $1.69 \times 10^{-4}$ & $4.76 \times 10^{-4}$ & $2.80 \times 10^{-4}$ \\
 $^{46}$Ca & $1.57 \times 10^{-18}$ & $1.54 \times 10^{-21}$ & $8.52 \times 10^{-24}$ & $4.65 \times 10^{-20}$ & $8.51 \times 10^{-22}$ & $4.60 \times 10^{-23}$ \\
 $^{48}$Ca & $1.32 \times 10^{-24}$ & $1.56 \times 10^{-24}$ & $2.55 \times 10^{-24}$ & $1.61 \times 10^{-24}$ & $2.30 \times 10^{-24}$ & $3.54 \times 10^{-24}$ \\
 $^{45}$Sc & $1.49 \times 10^{-6}$ & $6.45 \times 10^{-6}$ & $6.38 \times 10^{-8}$ & $1.17 \times 10^{-5}$ & $8.84 \times 10^{-6}$ & $5.12 \times 10^{-6}$ \\
 $^{46}$Ti & $1.00 \times 10^{-6}$ & $6.44 \times 10^{-6}$ & $8.6 \times 10^{-9}$ & $9.15 \times 10^{-6}$ & $2.85 \times 10^{-5}$ & $1.56 \times 10^{-6}$ \\
 $^{47}$Ti & $2.98 \times 10^{-6}$ & $4.17 \times 10^{-6}$ & $9.74 \times 10^{-7}$ & $1.28 \times 10^{-5}$ & $2.16 \times 10^{-5}$ & $7.60 \times 10^{-6}$ \\
 $^{48}$Ti & $1.75 \times 10^{-4}$ & $3.92 \times 10^{-4}$ & $1.54 \times 10^{-3}$ & $4.66 \times 10^{-4}$ & $1.47 \times 10^{-3}$ & $1.91 \times 10^{-3}$ \\
 $^{49}$Ti & $2.54 \times 10^{-6}$ & $5.31 \times 10^{-6}$ & $1.5 \times 10^{-6}$ & $3.11 \times 10^{-5}$ & $1.39 \times 10^{-5}$ & $5.69 \times 10^{-6}$ \\
 $^{50}$Ti & $7.50 \times 10^{-17}$ & $1.81 \times 10^{-19}$ & $6.93 \times 10^{-22}$ & $1.66 \times 10^{-18}$ & $2.69 \times 10^{-21}$ & $1.22 \times 10^{-22}$ \\
 $^{50}$V & $1.67 \times 10^{-13}$ & $6.27 \times 10^{-15}$ & $3.64 \times 10^{-15}$ & $8.35 \times 10^{-15}$ & $1.56 \times 10^{-16}$ & $1.33 \times 10^{-15}$ \\
 $^{51}$V & $3.89 \times 10^{-5}$ & $4.60 \times 10^{-5}$ & $1.84 \times 10^{-5}$ & $1.67 \times 10^{-4}$ & $3.85 \times 10^{-4}$ & $8.17 \times 10^{-5}$ \\
 $^{50}$Cr & $1.15 \times 10^{-5}$ & $1.00 \times 10^{-5}$ & $6.77 \times 10^{-7}$ & $6.50 \times 10^{-5}$ & $1.80 \times 10^{-4}$ & $8.57 \times 10^{-6}$ \\
 $^{52}$Cr & $2.24 \times 10^{-3}$ & $4.29 \times 10^{-3}$ & $1.62 \times 10^{-2}$ & $2.95 \times 10^{-3}$ & $1.12 \times 10^{-2}$ & $2.79 \times 10^{-2}$ \\
 $^{53}$Cr & $3.30 \times 10^{-5}$ & $2.53 \times 10^{-5}$ & $1.26 \times 10^{-4}$ & $3.80 \times 10^{-5}$ & $4.25 \times 10^{-5}$ & $1.87 \times 10^{-4}$ \\
 $^{54}$Cr & $1.20 \times 10^{-12}$ & $1.21 \times 10^{-13}$ & $1.67 \times 10^{-14}$ & $3.44 \times 10^{-14}$ & $1.99 \times 10^{-15}$ & $9.56 \times 10^{-15}$ \\
 $^{55}$Mn & $6.20 \times 10^{-5}$ & $7.82 \times 10^{-5}$ & $1.10 \times 10^{-4}$ & $1.52 \times 10^{-4}$ & $3.80 \times 10^{-4}$ & $3.48 \times 10^{-4}$ \\
 $^{54}$Fe & $1.67 \times 10^{-5}$ & $1.80 \times 10^{-5}$ & $2.88 \times 10^{-5}$ & $3.70 \times 10^{-5}$ & $5.27 \times 10^{-5}$ & $9.18 \times 10^{-5}$ \\
 $^{56}$Fe & $1.49 \times 10^{-1}$ & $2.47 \times 10^{-1}$ & $5.80 \times 10^{-1}$ & $2.60 \times 10^{-1}$ & $7.97 \times 10^{-1}$ & $11.84 \times 10^{-1}$ \\
 $^{57}$Fe & $1.58 \times 10^{-3}$ & $3.42 \times 10^{-3}$ & $7.32 \times 10^{-3}$ & $4.47 \times 10^{-3}$ & $1.31 \times 10^{-2}$ & $1.32 \times 10^{-2}$ \\
 $^{58}$Fe & $4.16 \times 10^{-13}$ & $3.70 \times 10^{-13}$ & $5.84 \times 10^{-13}$ & $5.59 \times 10^{-14}$ & $1.77 \times 10^{-13}$ & $8.44 \times 10^{-13}$ \\
 $^{60}$Fe & $1.93 \times 10^{-24}$ & $2.00 \times 10^{-25}$ & $2.87 \times 10^{-25}$ & $1.85 \times 10^{-25}$ & $2.73 \times 10^{-25}$ & $3.90 \times 10^{-25}$ \\
 $^{59}$Co & $6.87 \times 10^{-4}$ & $2.60 \times 10^{-4}$ & $2.8 \times 10^{-4}$ & $2.37 \times 10^{-3}$ & $3.59 \times 10^{-3}$ & $6.46 \times 10^{-4}$ \\
 $^{58}$Ni & $4.61 \times 10^{-4}$ & $3.68 \times 10^{-4}$ & $4.47 \times 10^{-4}$ & $1.89 \times 10^{-3}$ & $3.92 \times 10^{-3}$ & $1.38 \times 10^{-3}$ \\
 $^{60}$Ni & $4.30 \times 10^{-3}$ & $7.62 \times 10^{-3}$ & $1.27 \times 10^{-2}$ & $8.96 \times 10^{-3}$ & $2.32 \times 10^{-2}$ & $2.48 \times 10^{-2}$ \\
 $^{61}$Ni & $5.19 \times 10^{-5}$ & $1.17 \times 10^{-4}$ & $1.99 \times 10^{-4}$ & $1.80 \times 10^{-4}$ & $3.76 \times 10^{-4}$ & $3.78 \times 10^{-4}$ \\
 $^{62}$Ni & $2.80 \times 10^{-5}$ & $5.48 \times 10^{-5}$ & $1.28 \times 10^{-4}$ & $1.95 \times 10^{-4}$ & $3.12 \times 10^{-4}$ & $1.90 \times 10^{-4}$ \\
 $^{64}$Ni & $2.79 \times 10^{-18}$ & $9.67 \times 10^{-17}$ & $1.84 \times 10^{-15}$ & $6.42 \times 10^{-18}$ & $1.61 \times 10^{-17}$ & $3.19 \times 10^{-16}$ \\
 $^{63}$Cu & $2.10 \times 10^{-5}$ & $6.6 \times 10^{-6}$ & $1.00 \times 10^{-6}$ & $1.82 \times 10^{-4}$ & $1.00 \times 10^{-4}$ & $2.28 \times 10^{-5}$ \\
 $^{65}$Cu & $9.88 \times 10^{-7}$ & $2.79 \times 10^{-6}$ & $8.56 \times 10^{-6}$ & $3.80 \times 10^{-6}$ & $8.95 \times 10^{-6}$ & $1.20 \times 10^{-5}$ \\
 $^{64}$Zn & $2.35 \times 10^{-4}$ & $3.58 \times 10^{-4}$ & $7.10 \times 10^{-4}$ & $7.70 \times 10^{-4}$ & $1.82 \times 10^{-3}$ & $1.90 \times 10^{-3}$ \\
 $^{66}$Zn & $2.79 \times 10^{-6}$ & $5.50 \times 10^{-6}$ & $1.79 \times 10^{-5}$ & $1.87 \times 10^{-5}$ & $4.8 \times 10^{-5}$ & $2.22 \times 10^{-5}$ \\
 $^{67}$Zn & $4.58 \times 10^{-7}$ & $1.96 \times 10^{-7}$ & $1.44 \times 10^{-8}$ & $1.38 \times 10^{-6}$ & $7.98 \times 10^{-7}$ & $1.48 \times 10^{-6}$ \\
 $^{68}$Zn & $2.43 \times 10^{-6}$ & $5.50 \times 10^{-6}$ & $8.20 \times 10^{-6}$ & $4.90 \times 10^{-6}$ & $1.16 \times 10^{-5}$ & $1.19 \times 10^{-5}$ \\
 $^{70}$Zn & $1.21 \times 10^{-26}$ & $2.73 \times 10^{-26}$ & $3.53 \times 10^{-25}$ & $1.47 \times 10^{-26}$ & $3.98 \times 10^{-26}$ & $1.16 \times 10^{-25}$ \\

\end{longtable}

\begin{longtable}[c]{c c c c c c c}
 \caption{Same as Tab;e \ref{table:yield_isotopes} but for short-lived radioactive isotopes. \label{table:unstable_isotopes}} \\
 
 Isotope & $M=20, E=1$ & $M=25, E=1$ & $M=40, E=1$ & $M=20, E=10$ & $M=25, E=10$ & $M=40, E=10$\\
 \hline
 \endfirsthead

 \multicolumn{7}{c}{\textit{Continuation of Table \ref{table:unstable_isotopes}.}} \\
 Isotope & $M=20, E=1$ & $M=25, E=1$ & $M=40, E=1$ & $M=20, E=10$ & $M=25, E=10$ & $M=40, E=10$ \\
 \hline
 \endhead

 \hline
 \endfoot

 \hline
 \endlastfoot

 $^{22}$Na & $4.40 \times 10^{-7}$ & $7.56 \times 10^{-7}$ & $6.50 \times 10^{-8}$ & $3.99 \times 10^{-7}$ & $9.42 \times 10^{-7}$ & $1.37 \times 10^{-7}$ \\
 $^{26}$Al & $2.79 \times 10^{-6}$ & $3.33 \times 10^{-6}$ & $8.25 \times 10^{-7}$ & $5.34 \times 10^{-6}$ & $6.55 \times 10^{-6}$ & $1.90 \times 10^{-6}$ \\
 $^{39}$Ar & $8.79 \times 10^{-13}$ & $6.90 \times 10^{-14}$ & $7.90 \times 10^{-14}$ & $3.40 \times 10^{-13}$ & $1.55 \times 10^{-13}$ & $6.46 \times 10^{-14}$ \\
 $^{40}$K & $3.80 \times 10^{-11}$ & $8.55 \times 10^{-12}$ & $9.66 \times 10^{-12}$ & $8.61 \times 10^{-12}$ & $4.51 \times 10^{-12}$ & $4.26 \times 10^{-12}$ \\
 $^{41}$Ca & $8.42 \times 10^{-6}$ & $3.27 \times 10^{-6}$ & $1.46 \times 10^{-6}$ & $1.52 \times 10^{-6}$ & $1.14 \times 10^{-6}$ & $4.82 \times 10^{-6}$ \\
 $^{44}$Ti & $3.61 \times 10^{-5}$ & $8.20 \times 10^{-5}$ & $3.14 \times 10^{-4}$ & $1.67 \times 10^{-4}$ & $4.26 \times 10^{-4}$ & $2.85 \times 10^{-4}$ \\
 $^{48}$V & $2.23 \times 10^{-7}$ & $3.16 \times 10^{-8}$ & $2.68 \times 10^{-7}$ & $2.47 \times 10^{-9}$ & $1.54 \times 10^{-8}$ & $1.20 \times 10^{-7}$ \\
 $^{49}$V & $1.91 \times 10^{-7}$ & $1.30 \times 10^{-9}$ & $4.35 \times 10^{-9}$ & $5.41 \times 10^{-9}$ & $1.18 \times 10^{-9}$ & $2.68 \times 10^{-9}$ \\
 $^{53}$Mn & $1.14 \times 10^{-5}$ & $2.36 \times 10^{-7}$ & $3.52 \times 10^{-6}$ & $3.48 \times 10^{-8}$ & $8.13 \times 10^{-8}$ & $1.33 \times 10^{-6}$ \\
 $^{60}$Fe & $2.67 \times 10^{-23}$ & $5.16 \times 10^{-26}$ & $1.85 \times 10^{-25}$ & $6.65 \times 10^{-26}$ & $4.14 \times 10^{-26}$ & $1.29 \times 10^{-25}$ \\
 $^{56}$Co & $2.42 \times 10^{-5}$ & $2.85 \times 10^{-6}$ & $1.88 \times 10^{-5}$ & $1.86 \times 10^{-7}$ & $1.22 \times 10^{-6}$ & $1.19 \times 10^{-5}$ \\
 $^{57}$Co & $1.60 \times 10^{-6}$ & $1.32 \times 10^{-7}$ & $5.55 \times 10^{-7}$ & $2.62 \times 10^{-8}$ & $1.0 \times 10^{-7}$ & $4.64 \times 10^{-7}$ \\
 $^{60}$Co & $1.89 \times 10^{-18}$ & $6.87 \times 10^{-20}$ & $8.19 \times 10^{-20}$ & $2.85 \times 10^{-19}$ & $1.51 \times 10^{-20}$ & $5.85 \times 10^{-20}$ \\
 $^{56}$Ni & $1.49 \times 10^{-1}$ & $2.47 \times 10^{-1}$ & $5.80 \times 10^{-1}$ & $2.60 \times 10^{-1}$ & $7.97 \times 10^{-1}$ & $11.84 \times 10^{-1}$ \\
 $^{57}$Ni & $1.58 \times 10^{-3}$ & $3.41 \times 10^{-3}$ & $7.33 \times 10^{-3}$ & $4.45 \times 10^{-3}$ & $1.30 \times 10^{-2}$ & $1.33 \times 10^{-2}$ \\
 $^{59}$Ni & $5.54 \times 10^{-4}$ & $9.28 \times 10^{-6}$ & $1.71 \times 10^{-5}$ & $2.89 \times 10^{-5}$ & $3.34 \times 10^{-5}$ & $3.31 \times 10^{-5}$ \\
 $^{63}$Ni & $2.70 \times 10^{-19}$ & $4.81 \times 10^{-19}$ & $2.42 \times 10^{-18}$ & $8.6 \times 10^{-20}$ & $1.54 \times 10^{-19}$ & $1.53 \times 10^{-18}$ \\

 \end{longtable}

 \begin{longtable}[c]{c c c c c c c}
 \caption{Same as \ref{table:yield_isotopes} but for the elemental yields. \label{table:yield_elements}} \\
 
 Isotope & $M=20, E=1$ & $M=25, E=1$ & $M=40, E=1$ & $M=20, E=10$ & $M=25, E=10$ & $M=40, E=10$ \\
 \hline
 \endfirsthead

 \multicolumn{7}{c}{\textit{Continuation of Table \ref{table:yield_elements}.}} \\
 Isotope & $M=20, E=1$ & $M=25, E=1$ & $M=40, E=1$ & $M=20, E=10$ & $M=25, E=10$ & $M=40, E=10$ \\
 \hline
 \endhead

 \hline
 \endfoot

 \hline
 \endlastfoot

 C & $2.12 \times 10^{-1}$ & $3.6 \times 10^{-1}$ & $4.33 \times 10^{-1}$ & $1.89 \times 10^{-1}$ & $2.73 \times 10^{-1}$ & $4.32 \times 10^{-1}$ \\
 N & $1.10 \times 10^{-6}$ & $4.29 \times 10^{-6}$ & $9.57 \times 10^{-6}$ & $1.29 \times 10^{-6}$ & $6.23 \times 10^{-6}$ & $1.21 \times 10^{-5}$ \\
 O & 2.15 & 2.84 & 8.24 & 2.02 & 2.26 & 7.50 \\
 F & $1.89 \times 10^{-10}$ & $1.93 \times 10^{-10}$ & $8.97 \times 10^{-11}$ & $1.61 \times 10^{-10}$ & $6.15 \times 10^{-10}$ & $2.66 \times 10^{-10}$ \\
 Ne & $9.19 \times 10^{-1}$ & $5.18 \times 10^{-1}$ & $3.11 \times 10^{-1}$ & $7.29 \times 10^{-1}$ & $2.53 \times 10^{-1}$ & $3.25 \times 10^{-1}$ \\
 Na & $2.22 \times 10^{-5}$ & $2.90 \times 10^{-5}$ & $1.40 \times 10^{-6}$ & $3.56 \times 10^{-5}$ & $5.64 \times 10^{-5}$ & $5.65 \times 10^{-6}$ \\
 Mg & $1.45 \times 10^{-1}$ & $1.10 \times 10^{-1}$ & $4.63 \times 10^{-1}$ & $1.39 \times 10^{-1}$ & $9.12 \times 10^{-2}$ & $3.89 \times 10^{-1}$ \\
 Al & $8.95 \times 10^{-5}$ & $1.79 \times 10^{-4}$ & $1.10 \times 10^{-4}$ & $2.20 \times 10^{-4}$ & $3.38 \times 10^{-4}$ & $1.76 \times 10^{-4}$ \\
 Si & $9.20 \times 10^{-2}$ & $3.90 \times 10^{-1}$ & $9.22 \times 10^{-1}$ & $1.46 \times 10^{-1}$ & $4.15 \times 10^{-1}$ & $9.71 \times 10^{-1}$ \\
 P & $3.84 \times 10^{-5}$ & $1.00 \times 10^{-4}$ & $1.27 \times 10^{-4}$ & $1.14 \times 10^{-4}$ & $1.85 \times 10^{-4}$ & $2.64 \times 10^{-4}$ \\
 S & $5.68 \times 10^{-2}$ & $1.76 \times 10^{-1}$ & $4.85 \times 10^{-1}$ & $6.60 \times 10^{-2}$ & $2.32 \times 10^{-1}$ & $5.31 \times 10^{-1}$ \\
 Cl & $3.77 \times 10^{-5}$ & $4.35 \times 10^{-5}$ & $6.17 \times 10^{-5}$ & $1.17 \times 10^{-4}$ & $1.93 \times 10^{-4}$ & $1.60 \times 10^{-4}$ \\
 Ar & $1.26 \times 10^{-2}$ & $3.56 \times 10^{-2}$ & $9.83 \times 10^{-2}$ & $1.38 \times 10^{-2}$ & $5.16 \times 10^{-2}$ & $1.90 \times 10^{-1}$ \\
 K & $2.52 \times 10^{-5}$ & $3.53 \times 10^{-5}$ & $3.3 \times 10^{-5}$ & $7.59 \times 10^{-5}$ & $1.31 \times 10^{-4}$ & $7.79 \times 10^{-5}$ \\
 Ca & $1.13 \times 10^{-2}$ & $3.34 \times 10^{-2}$ & $1.0 \times 10^{-1}$ & $1.47 \times 10^{-2}$ & $5.60 \times 10^{-2}$ & $1.11 \times 10^{-1}$ \\
 Sc & $1.49 \times 10^{-6}$ & $6.45 \times 10^{-6}$ & $6.38 \times 10^{-8}$ & $1.17 \times 10^{-5}$ & $8.84 \times 10^{-6}$ & $5.12 \times 10^{-6}$ \\
 Ti & $1.81 \times 10^{-4}$ & $4.80 \times 10^{-4}$ & $1.54 \times 10^{-3}$ & $5.19 \times 10^{-4}$ & $1.53 \times 10^{-3}$ & $1.93 \times 10^{-3}$ \\
 V & $3.89 \times 10^{-5}$ & $4.60 \times 10^{-5}$ & $1.84 \times 10^{-5}$ & $1.67 \times 10^{-4}$ & $3.85 \times 10^{-4}$ & $8.17 \times 10^{-5}$ \\
 Cr & $2.28 \times 10^{-3}$ & $4.32 \times 10^{-3}$ & $1.64 \times 10^{-2}$ & $3.50 \times 10^{-3}$ & $1.13 \times 10^{-2}$ & $2.81 \times 10^{-2}$ \\
 Mn & $6.20 \times 10^{-5}$ & $7.82 \times 10^{-5}$ & $1.10 \times 10^{-4}$ & $1.52 \times 10^{-4}$ & $3.80 \times 10^{-4}$ & $3.48 \times 10^{-4}$ \\
 Fe & $1.51 \times 10^{-1}$ & $2.51 \times 10^{-1}$ & $5.87 \times 10^{-1}$ & $2.64 \times 10^{-1}$ & $8.10 \times 10^{-1}$ & $11.97 \times 10^{-1}$ \\
 Co & $6.87 \times 10^{-4}$ & $2.60 \times 10^{-4}$ & $2.80 \times 10^{-4}$ & $2.37 \times 10^{-3}$ & $3.59 \times 10^{-3}$ & $6.46 \times 10^{-4}$ \\
 Ni & $4.84 \times 10^{-3}$ & $8.16 \times 10^{-3}$ & $1.35 \times 10^{-2}$ & $1.11 \times 10^{-2}$ & $2.79 \times 10^{-2}$ & $2.68 \times 10^{-2}$ \\
 Cu & $2.20 \times 10^{-5}$ & $8.85 \times 10^{-6}$ & $9.56 \times 10^{-6}$ & $1.85 \times 10^{-4}$ & $1.90 \times 10^{-4}$ & $3.31 \times 10^{-5}$ \\
 Zn & $2.41 \times 10^{-4}$ & $3.69 \times 10^{-4}$ & $7.36 \times 10^{-4}$ & $7.94 \times 10^{-4}$ & $1.87 \times 10^{-3}$ & $1.12 \times 10^{-3}$ \\

\end{longtable}

\vfill

\bibliographystyle{aasjournal}
\pagestyle{plain}
\bibliography{biblio}

\end{document}